\documentclass[article,showpacs,amsmath,amssymb,superscriptaddress]{revtex4}
\input{psfig.sty}
\begin{document}
\draft

\title{Semiclassical kinetic theory of electron spin relaxation in semiconductors}

\author{Franz X.~Bronold}
\affiliation{Institut f\"ur Physik,
Ernst-Moritz-Arndt-Universit\"at Greifswald,
D-17487 Greifswald,  Germany}
\author{Avadh Saxena}
\affiliation{Theoretical Division, Los Alamos National Laboratory,
Los Alamos, New Mexico 87545}
\author{Darryl L. Smith}
\affiliation{Theoretical Division, Los Alamos National Laboratory,
Los Alamos, New Mexico 87545}
\date{\today}

\begin{abstract}
{We develop a semiclassical kinetic theory for electron spin
relaxation in semiconductors. Our approach accounts for elastic
as well as inelastic scattering and treats Elliott-Yafet and
motional-narrowing processes, such as D'yakonov-Perel' and variable
g-factor processes, on an equal footing. Focusing on small spin
polarizations and small momentum transfer scattering, we derive,
starting from the full quantum kinetic equations, a Fokker-Planck
equation for the electron spin polarization. We then construct,
using a rigorous multiple time scale approach, a Bloch
equation for the macroscopic ($\vec{k}$-averaged) spin polarization
on the long time scale, where the spin polarization decays.
Spin-conserving energy relaxation and diffusion, which occur
on a fast time scale, after the initial spin polarization
has been injected, are incorporated and shown to give rise to
a weight function which defines the energy averages required for
the calculation of the spin relaxation tensor in the Bloch
equation. Our approach provides an intuitive way to conceptualize
the dynamics of the spin polarization in terms of a  ``test''
spin polarization which scatters off ``field'' particles
(electrons, impurities, phonons). To illustrate our approach, we
calculate for a quantum well the spin lifetime at temperatures
and densities where electron-electron and electron-impurity
scattering dominate. The spin lifetimes are non-monotonic
functions of temperature and density. Our results show that at
electron densities and temperatures, where the cross-over
from the non-degenerate to the degenerate regime occurs, spin
lifetimes are particularly long.
}
\end{abstract}


\maketitle

\section{Introduction}

The spin degree of freedom of an electron provides an
additional variable that potentially can be
used to add new functionality to electronic, optoelectronic and
magnetoelectronic devices or to even build radically new devices
entirely based on the coherence of electron spin states. This has
led to the newly emerging field of spintronics.~\cite{Prinz,Wolf01}
A subclass of spintronics device concepts relies on the capability to
inject, control, and detect electron spin polarizations in non-magnetic
semiconductors.~\cite{springer,semicon} The spin polarization, which
would enable the device operation, is a non-equilibrium state and its
characterization, e.g., in terms of lifetimes and transport coefficients,
has to be given within a kinetic theory.

Of particular interest is the lifetime of the non-equilibrium spin
polarization in non-magnetic n-type III-V semiconductors. Important
spin relaxation processes for itinerant electrons in this class of
materials include the Elliott-Yafet (EY) process~\cite{Elliott54,Yafet63},
that leads to spin-flip scattering and, in materials without
inversion symmetry, the D'yakonov-Perel' (DP) process~\cite{Dyakonov72}
in which spin states precess because of spin off-diagonal Hamiltonian
matrix elements resulting from a combination of spin-orbit
coupling and inversion asymmetry. An external magnetic field, in
many cases required to control and manipulate the electron spin,
can also influence the electron spin dynamics. It quenches
the DP process~\cite{Ivchenko73}, thereby tending to extend the
spin lifetimes as a function of magnetic field, and it opens
a spin relaxation channel due to the $\vec{k}$-dependence of the electron
g-factor, which forces the spin of electrons in different quantum
states to precess around an external magnetic field with different
rates.~\cite{Margulis83,our} For brevity we will refer to this mechanism
as a variable $g$-factor (VG) process.

Spin dynamics in semiconductors has been extensively studied in
magneto-optics~\cite{Chazalviel75,Fishman77,Ioffe,OptOrientation}
using various spin-sensitive emission, transmission, and reflection
spectroscopies. These spectroscopies are now readily adaptable to
spatially and time resolved
measurements~\cite{Vina99,scott,Kikkawa97,Kikkawa98,Kikkawa99,Poel89,Damen91,Srinivas93,Wagner93,Terauchi99,Tackeuchi99,Ohno99,Triques99,Boggess00,Malinowski00,Sandhu01,Olesberg01,Dzhioev02a,Dzhioev02b} which, together with the emergence of spintronics concepts, inspired
new theoretical investigations in
bulk~\cite{our,Wu00a,Wu01,Song02,Semenov03,Glazov03a} and dimension-reduced
semiconductors.~\cite{SIA,Dyakonov86,Bastard92,Sham93,Averkiev99,WuMetiu00,Lau01,Glazov02,Brand02,Glazov03b,Kainz03,Puller03,weng1,weng2,weng3}

The theoretical investigations are based on
the early work~\cite{Elliott54,Yafet63,Dyakonov72} augmented by
modern band structure theory for bulk and dimension-reduced
semiconductors. The EY spin relaxation rates are
usually calculated using the Golden rule for spin-flip
scattering, whereas the spin-flip rates due to motional-narrowing
(DP and VG) processes are, at least conceptually, obtained
from a semiclassical Boltzmann-type equation for the non-equilibrium
spin polarization, although not always is the Boltzmann equation explicitly
solved. Instead, a common procedure is to adapt the expression for
the spin relaxation rate originally derived by D'yakonov and
Perel'~\cite{Dyakonov72} to the scattering processes under consideration.

The D'yakonov-Perel' expression for the spin relaxation rate, which 
results from the solution of the Boltzmann equation within the
elastic approximation, treats all scattering processes on-shell,
even inelastic scattering events, e.g., due to electron-electron or
electron-phonon scattering.
The obtained spin relaxation rates are therefore on-shell rates, which have
to be averaged over energy with an appropriate weight function before a
comparison with experiments can be attempted. Usually, the
difference of the distribution functions for spin-up and spin-down
electrons divided by the total number of electrons contributing to the
initial spin polarization is used as a weight
function.~\cite{Dyakonov72,Dyakonov86,Averkiev99} The ad-hoc
energy averaging, which is necessary because of the incomplete
treatment of inelastic scattering processes, can lead to substantial
deviations from the spin relaxation rates obtained, e.g., from a full
numerical solution of the Boltzmann equation.~\cite{Wu01}

We develop in this paper a systematic kinetic theory for electron
spin relaxation, applicable to spin-flip (EY) and motional-narrowing
(DP and VG) spin relaxation processes in bulk and quantum wells, which
avoids the ad-hoc energy averaging and gives a clear physical picture of
the time evolution of the optically or electrically injected
non-equilibrium spin polarization. We derive, starting
from the full quantum kinetic equations for the electron Green functions,
a semiclassical Fokker-Planck equation for the time evolution of
the non-equilibrium spin polarization, valid for small spin polarizations
and for small momentum transfer scattering, and employ a multiple
time scale perturbation approach to separate the fast spin-conserving
from the slow spin non-conserving time evolution. As a result,
we obtain on the time scale of spin relaxation a Bloch equation for
the macroscopic ($\vec{k}$-averaged) non-equilibrium
spin polarization, which is the quantity measured
in, e.g., time-resolved Faraday and Kerr rotation
experiments.~\cite{scott,Kikkawa97,Kikkawa98,Kikkawa99,Malinowski00,Sandhu01}
The weight function defining the energy averages needed, e.g., for the
calculation of the spin relaxation tensor and the spin relaxation
rates, turns out to be directly related to the quasi-stationary spin
polarization, which is the terminating state of the initial, fast
spin-conserving time evolution of the injected spin polarization.
Our approach treats spin-flip (EY) and motional-narrowing (DP and VG)
processes on an equal footing. Due to the different angle dependences,
a Matthiessen-type rule holds, however, for isotropic
semiconductors, where the total spin relaxation tensor
is simply the sum of the individual spin relaxation tensors.~\cite{our}
The diagonal elements of the spin relaxation tensor, the
spin relaxation rates, are either given in terms of an energy
averaged spin-flip rate (EY process) or an energy
average of a generalized relaxation time, which accounts for both on-
and off-shell scattering events, multiplied by a precession rate
(DP and VG processes).

In the next section we introduce a generic model for
electrons in n-type III-V semiconductors applicable to bulk and
quantum well situations.
In Sec. III we give a complete description of our semiclassical kinetic
theory for the electron spin dynamics. As far as the
formal development is concerned, we treat EY, DP and VG processes
on an equal footing and also allow for quenching effects due
to orbital motion of electrons in an external magnetic
field. In Sec. IV we apply our approach to the particular
situation of DP spin relaxation in
an idealized quantum well at temperatures and densities for
which electron-electron and electron-impurity scattering dominate.
Our main findings are summarized in Sec. V. Technical details
concerning the calculation of the quantum well collision integrals
due to electron-electron and electron-impurity scattering are
relegated to an Appendix.

\section{Model Hamiltonian}

We consider conduction band (CB) electrons in III-V semiconductors,
e.g., GaAs, in the presence of an applied magnetic field. The model
used here applies to both bulk and quantum well situations. Within
an envelope function approach\cite{Ogg66,Golubev85}, which treats
the two states at the conduction band minimum explicitly and
includes a large set of states perturbatively, the
effective mass Hamiltonian for the CB electrons
can be cast into the form
\begin{eqnarray}
  {H}_{\alpha\alpha'}(\vec{K})=
  \epsilon(\vec{K})\delta_{\alpha\alpha'}
  +{\hbar\over 2}
  \vec{\Omega}_L\!\cdot\!\vec{\sigma}_{\alpha\alpha'}
  +{\hbar\over 2}[\vec{\Omega}_{IA}(\vec{K})+
  \vec{\Omega}_g(\vec{K})]
  \!\cdot\!\vec{\sigma}_{\alpha\alpha'},
\label{model}
\end{eqnarray}
where $\vec{K}=\vec{k}-(e/\hbar c)\vec{A}(\vec{r})$ and
$\vec{A}(\vec{r})$ is the vector potential. The ``spin basis'' for
the CB electrons used to define the model (\ref{model}) is
$\alpha=+$ and $\alpha=-$, where $\alpha=+$ ($\alpha=-$) denotes a
state which is mostly spin-up (spin-down) with a small admixture
of spin-down (spin-up).

The first term denotes the dispersion of the Kramers degenerate
conduction band which, depending on the sophistication of the
envelope function approach, could contain nonparabolicity effects.
For quantum wells $\vec{k}$ and $\epsilon(\vec{k})$ denote the
in-plane momentum and the in-plane dispersion of the conduction
subband under consideration.  The second term comprises the Larmor
precession due to the external magnetic field, with $\hbar\vec
{\Omega}_L=\mu_B g^*\vec{B}$ the Larmor energy vector. Here $\mu_B$
and $g^*$ denote the Bohr magneton and the electron $g$-factor. The
third term describes spin off-diagonal
Hamiltonian matrix elements arising from the coupling to higher
lying states. The most important of which are
the splitting of the conduction band due to inversion asymmetry (IA)
and the term which leads to a $\vec{k}$-dependent electron
g-factor. For bulk semiconductors, the two contributions are given by
\begin{eqnarray}
\hbar\vec{\Omega}_{IA}(\vec{K})&=&2\delta_0\vec{\kappa}_{IA}(\vec{K})~,
\label{IA}\\
\hbar\vec{\Omega}_g(\vec{K})&=&2 a_4 K^2 \vec{B}
  + 2 a_5\{\vec{K},\vec{B}\!\cdot\!\vec{K}\}
  + 2 a_6\vec{\tau}(\vec{K},\vec{B})~,
\label{VG}
\end{eqnarray}
respectively. The definition of the vectors $\vec{\kappa}_{IA}(\vec{K})$
and $\vec{\tau}(\vec{K},\vec{B})$ and of the parameters $\delta_0$ and
$a_i$ can be found in Refs.~\cite{Ogg66,Golubev85} and $\{..,..\}$
denotes an anticommutator. The expressions for conduction subbands
in a quantum well are obtained by averaging the bulk expressions
(\ref{IA}) and (\ref{VG}) over the subband envelope function.

In addition to bulk inversion asymmetry, dimension-reduced
semiconductors can have additional sources of asymmetry due to
interfaces which share no common atom~\cite{Olesberg01} or due to
layer design (structural inversion asymmetry~\cite{SIA}). Both
mechanisms can be cast into spin off-diagonal Hamiltonian matrix
elements and can be therefore treated in the same way as the spin
off-diagonal terms due to bulk inversion asymmetry.

For a complete description, a collision term arising from
electron-impurity, electron-phonon and electron-electron scattering,
$$H_c=H_{ei}+H_{ep}+H_{ee}~,$$
is added to the effective mass Hamiltonian.
The electron-impurity term reads
\begin{eqnarray}
H_{ei}=\sum_{\vec{k}\vec{k'}}\sum_{\alpha\alpha'}
M_{\alpha,\alpha'}(\vec{k},\vec{k}')
c^\dag_{\vec{k}\alpha}c_{\vec{k}'\alpha'}~,
\label{ei}
\end{eqnarray}
with a scattering matrix element given by
\begin{eqnarray}
M_{\alpha,\alpha'}(\vec{k},\vec{k}')=\sum_j
U(\vec{k}-\vec{k'})
e^{i(\vec{k}-\vec{k'})\vec{r}_j} I_{\alpha\alpha'}(\vec{k},\vec{k}')~.
\label{Mei}
\end{eqnarray}
The Bloch states for the conduction band
are not pure spin states, because of spin-orbit coupling. The scattering matrix
element contains therefore
an overlap factor
\begin{eqnarray}
I_{\alpha\alpha'}(\vec{k},\vec{k}')=
\langle U_{\alpha,\vec k} |U_{\alpha',\vec{k}'} \rangle~,
\label{overlap}
\end{eqnarray}
which is of order unity for $\alpha=\alpha'$ (spin conserving scattering) and is
small, but not zero otherwise (spin non-conserving scattering). The
electron-phonon collision term would have the same structure as Eq. (\ref{ei})
but with phonon creation and annihilation operators appearing in the
matrix element $M_{\alpha\alpha'}(\vec{k},\vec{k}')$.
The electron-electron scattering contribution has the form
\begin{eqnarray}
H_{ee}=\frac{1}{2}\sum_{\vec{k}_i\alpha_i}
M_{\alpha_1\alpha_2\alpha_3\alpha_4}(\vec{k}_1,\vec{k}_2,\vec{k}_3,\vec{k}_4)
c^\dag_{\vec{k}_1\alpha_1}c^\dag_{\vec{k}_2\alpha_2}
c_{\vec{k}_3\alpha_3}c_{\vec{k}_4\alpha_4}~,
\label{ee}
\end{eqnarray}
where the scattering matrix element,
\begin{eqnarray}
M_{\alpha_1\alpha_2\alpha_3\alpha_4}
(\vec{k}_1,\vec{k}_2,\vec{k}_3,\vec{k}_4)=
V(\vec{k}_1-\vec{k}_4)
I_{\alpha_1\alpha_4}(\vec{k}_1,\vec{k}_4)I_{\alpha_2\alpha3}
(\vec{k}_2,\vec{k}_3)\delta_{\vec{k}_1+\vec{k}_2,\vec{k}_3+\vec{k}_4}~,
\label{Mee}
\end{eqnarray}
contains two overlap factors. The functions $U(\vec{k})$ and $V(\vec{k})$
denote, respectively, the potential of a single impurity (neutral or ionized)
and the Coulomb potential between two conduction electrons.

The model Hamiltonian is characterized by
$\epsilon(\vec{k})$, $\vec{\Omega}_{IA}(\vec{k})$, $\vec{\Omega}_{g}(\vec{k})$,
and $I_{\alpha,\alpha'}(\vec{k},\vec{k}')$. These quantities need to be
obtained by an electronic structure calculation. The formal structure of
the kinetic theory described in the next section is independent
of the particular form of these quantities.

\section{Semiclassical kinetic theory}

In this section we give a systematic derivation of the Fokker-Planck
equation governing the electron spin relaxation in the limit of
small spin polarizations. The derivation is independent of
dimensionality, applying to bulk semiconductors and
semiconductor heterostructures, and treats motional-narrowing
(DP and VG) and spin-flip (EY) spin relaxation processes
on an equal footing.
To obtain a Fokker-Planck equation, we restrict ourselves to the
Born approximation, but collective effects giving rise to dynamical
screening of the Coulomb interaction can be approximately incorporated
at the level of a quantum analog to the Lenard-Balescu equation.~\cite{LB}
Besides its
intuitive interpretation in terms of a small ``test'' spin polarization
scattering off a bath of ``field'' particles (impurities, electrons and
phonons), causing dynamical friction, diffusion, and eventually
relaxation for the ``test'' spin polarization,
the Fokker-Planck equation is the starting point for a multiple
time scale analysis which results in the derivation of a Bloch
equation for the macroscopic ($\vec{k}$-averaged) spin polarization.
Its decay is  usually characterized by the diagonal elements of a
spin relaxation tensor, which are quadratures of either a spin-flip
rate (EY process) or a generalized relaxation time
multiplied by a precession rate (DP and VG processes).

Since the derivation is quite lengthy and to some extent rather formal
we first give a short outline of the main steps. We start from the
full quantum kinetic equations for the Keldysh Green
functions.~\cite{Keldysh65,Kinetic,Kuznetsov91,HaugJauho} Each component
of the Keldysh Green function is a $2\times 2$ matrix in electron spin
space. In the first step we derive,
within the semiclassical approximation, a kinetic equation
for the density matrix. This accounts to treating momentum
scattering processes as instantaneous on the time scale of spin
relaxation, which is usually the case. Calculating the self-energies
which appear in the semiclassical kinetic equation in the Born
approximation, linearizing with respect to spin polarization,
and expanding the self-energies up to second order in the momentum
transfer (diffusion approximation) finally yields a Fokker-Planck
equation for the spin polarization, which we then analyze in terms
of multiple time scale perturbation theory.

\subsection{Kinetic equations}

For a spatially homogeneous system (we assume a constant magnetic
field $\vec{B}$), the information about spin relaxation is contained in
the electronic density matrix, which, due to the spin degree
of freedom, is a $2\times 2$ matrix in spin space,
\begin{eqnarray}
N_{\alpha_1\alpha_2}(\vec{k},t)=
\large<(c_{\vec{k}\alpha_1}^\dagger c_{\vec{k}\alpha_2})(t)
\large>~,
\label{DM}
\end{eqnarray}
but diagonal in $\vec{k}$-space. Here, the operators evolve in
time with the full Hamiltonian, including the time-dependent
perturbation, which could be, e.g., a circularly polarized light
pulse applied at time $t=t_0$. To perform the averaging in Eq.
(\ref{DM}) denoted by $\large<[...]\large>$, we
consider the system to be in thermodynamical equilibrium for
$t<t_0$, take the limit $t_0\rightarrow -\infty$ and evaluate the
expectation value in Eq. (\ref{DM}) with respect to the
equilibrium density matrix.~\cite{HaugJauho}

To derive a kinetic equation for the density matrix it is
convenient to start from Keldysh Green functions.~\cite{Keldysh65,Kinetic}
For a constant magnetic field, the vector potential is a function of
$\vec{r}$. It is therefore necessary to initially work with kinetic
equations in real space. In this subsection we set $\hbar=1$.
Introducing a numerical index $1$ that stands for $\vec{r}_1\alpha_1 t_1$
and $2$ for $\vec{r}_2\alpha_2 t_2$, we write in the notation of
Ref.~\cite{Kuznetsov91}
\begin{eqnarray}
i\hat{G}_{12}=i\left( \begin{array}{cc}
G_{12}^{++} & G_{12}^{+-}\\
G_{12}^{-+} & G_{12}^{--}
\end{array}
\right) .
\end{eqnarray}
Note that each component of the Keldysh Green function is a
$2\times 2$ matrix in spin space. Introducing further a self-energy
\begin{eqnarray}
\hat{\Sigma}_{12}=\left( \begin{array}{cc}
\Sigma_{12}^{++} & \Sigma_{12}^{+-}\\
\Sigma_{12}^{-+} & \Sigma_{12}^{--}
\end{array}
\right),
\end{eqnarray}
we set up two matrix Dyson equations, one where the time differentiation
is with respect to $t_1$ and one where it is with respect to $t_2$:
\begin{eqnarray}
\partial_{t_1}\hat{G}_{12}&=&-i\hat{\tau}_z\delta_{12}
-i(\hat{\epsilon}\hat{G})_{12}-i\hat{\tau}_z(\hat{\Sigma}\hat{G})_{12}~,
\label{Dyson1}\\
\partial_{t_2}\hat{G}_{12}&=&i\hat{\tau}_z\delta_{12}
+i(\hat{G}\hat{\epsilon})_{12}+i(\hat{G}\hat{\Sigma})_{12}\hat{\tau}_z~,
\label{Dyson2}
\end{eqnarray}
with $\delta_{12}=\delta(t_1-t_2)\delta(12)$ and the energy matrix
$\hat{\epsilon}_{12}=\delta(t_1-t_2)
\hat{\epsilon}(12)=\delta(t_1-t_2)
\hat{\tau}_0\delta(12)\epsilon(-i\nabla_{\vec{r}_1}
-(e/c)\vec{A}(\vec{r}_1))$, where we neglect nonparabolicities in the dispersion.
We adopt the convention that numerical indices written as a subscript contain
the time variable, whereas numerical indices written as an
argument do not. Matrix multiplication with respect to the Keldysh indices
is implied and internal variables are summed (integrated) over; 
$\hat{\tau}_z$ is a Pauli matrix and $\hat{\tau}_0$ is the unit matrix in
Keldysh space.  

Subtracting Eq. (\ref{Dyson2}) from Eq. (\ref{Dyson1}) gives
\begin{eqnarray}
\large[\hat{L},\hat{G}\large]_{12}=
\hat{\tau}_z(\hat{\Sigma}\hat{G})_{12}
-(\hat{G}\hat{\Sigma})_{12}\hat{\tau}_z~,
\label{Dyson3}
\end{eqnarray}
where $[..,..]$ denotes the commutator. To compactify the notation, we introduced
a differential operator
\begin{eqnarray}
\hat{L}_{13} = \hat{\tau}_0\delta_{13} L(3) =
\hat{\tau}_0 \delta_{13}
\large(i\partial_{t_3}-\epsilon(3)\large)~.
\end{eqnarray}
It is understood that in the second term of the commutator, the operator
$\hat{L}_{32}$ acts to the left with the temporal differential
operator $\partial_{t_3}$ replaced by its adjoint $-\partial_{t_3}$.

Equation (\ref{Dyson3}) contains two time variables. To obtain a kinetic
equation for the electronic density matrix, which depends only on a single
time variable, it is
necessary to perform the equal time limit. This is most conveniently done
in the (mixed) Wigner representation, where the equal time limit reduces to
an integration. 
Separating the self-energy into a singular and a regular part~\cite{Kuznetsov91}
\begin{eqnarray}
\Sigma^{pq}_{12}=\Delta^{pq}(12;t_1)\delta_{pq}\delta(t_1-t_2)+\tilde{\Sigma}^{pq}_{12}~,
\end{eqnarray}   
introducing relative and center variables,
$\vec{r}=\vec{r}_1-\vec{r}_2$, $\vec{R}=(\vec{r}_1+\vec{r}_2)/2$,
$\tau=t_1-t_2$, and $T=(t_1+t_2)/2$, and defining a Fourier transformation
with respect to the relative variables, 
\begin{eqnarray}
A(\vec{R},T,\vec{k},\omega)=
\int_{-\infty}^\infty d\tau \int d\vec{r} e^{i\omega\tau-i\vec{k}\vec{r}}
A(\vec{R},T,\vec{r},\tau)~,
\end{eqnarray}
together with a gradient operator~\cite{HaugJauho}
\begin{eqnarray}
{\cal G}^{AB}=
\exp{{1\over{2i}}
\left[\partial_T^A\partial_\omega^B-\partial_\omega^A\partial_T^B
+\nabla_{\vec{k}}^A \cdot \nabla_{\vec{R}}^B
-\nabla_{\vec{R}}^A \cdot \nabla_{\vec{k}}^B\right]}~,
\end{eqnarray}          
the equal time limit of the $++$ component of Eq. (\ref{Dyson3}) can be
written as
\begin{eqnarray}
D(\vec{R},T,\vec{k})=F(\vec{R},T,\vec{k})+C(\vec{R},T,\vec{k})~,   
\label{Dyson4}
\end{eqnarray}  
with a driving term on the lhs, 
\begin{eqnarray}  
D(\vec{R},T,\vec{k})&=&\int_{-\infty}^\infty {{d\omega}\over{2\pi}}
\large( {\cal G}^{LG}  L(\vec{R},T,\vec{k},\omega) 
                       G^{++}(\vec{R},T,\vec{k},\omega)
\nonumber\\
      &-& {\cal G}^{GL}  G^{++}(\vec{R},T,\vec{k},\omega)
                       L(\vec{R},T,\vec{k},\omega) \large)~,
\label{driving1}
\end{eqnarray}
and a rhs, which contains a molecular field term arising from the 
singular part of the self-energy, 
\begin{eqnarray}
F(\vec{R},T,\vec{k})&=&
\int_{-\infty}^{\infty}{{d\omega}\over{2\pi}}
\large[ {\cal G}^{\Delta G} \Delta^{++}(\vec{R},T,\vec{k})
                                G^{++}(\vec{R},T,\vec{k},\omega)
\nonumber\\
      &-& {\cal G}^{G\Delta}  G^{++}(\vec{R},T,\vec{k},\omega)
                       \Delta^{++}(\vec{R},T,\vec{k}) \large]~,   
\label{molecular}
\end{eqnarray}
and a collision term due to the regular part 
\begin{eqnarray}
C(\vec{R},T,\vec{k})&=&
\int_{-\infty}^{\infty}{{d\omega}\over{2\pi}}
\large( {\cal G}^{\tilde{\Sigma}G} 
                 \tilde{\Sigma}^{++}(\vec{R},T,\vec{k},\omega)
                              G^{++}(\vec{R},T,\vec{k},\omega)
\nonumber\\
      &-& {\cal G}^{G\tilde{\Sigma}} G^{++}(\vec{R},T,\vec{k},\omega)
                       \tilde{\Sigma}^{++}(\vec{R},T,\vec{k},\omega) 
\nonumber\\
&+&{\cal G}^{\tilde{\Sigma}G}
\tilde{\Sigma}^{+-}(\vec{R},T,\vec{k},\omega)G^{-+}(\vec{R},T,\vec{k},\omega)
\nonumber\\
&-&{\cal G}^{G\tilde{\Sigma}}
G^{+-}(\vec{R},T,\vec{k},\omega)\tilde{\Sigma}^{-+}(\vec{R},T,\vec{k},\omega) 
\large)~.
\label{collision}
\end{eqnarray}

The semiclassical approximation amounts to the assumption that the
Green functions and self-energies vary slowly on the macroscopic
scales, $T$ and $\vec{R}$, respectively. It is therefore sufficient 
to keep in Eq. ({\ref{Dyson4}) only the leading order terms in a 
gradient expansion. The leading order of the rhs of Eq. ({\ref{Dyson4}) 
is the zeroth order, i.e., Eqs. (\ref{molecular}) and (\ref{collision}) 
with ${\cal G}^{AB}\rightarrow 1$. The lhs of Eq. ({\ref{Dyson4}), however,
has to be determined to first order, because the zeroth order vanishes.
Using 
\begin{eqnarray}
L(\vec{R},T,\omega,\vec{k})=\omega+\epsilon(\vec{K})~,
\end{eqnarray} 
we explicitly obtain   
\begin{eqnarray}
D(\vec{R},T,\vec{k})&=&
i\int_{-\infty}^{\infty}{{d\omega}\over{2\pi}}
[\partial_T G^{++}(\vec{R},T,\vec{k},\omega)
- \nabla_{\vec{R}}~\epsilon(\vec{K})\cdot\nabla_{\vec{k}}~ 
G^{++}(\vec{R},T,\vec{k},\omega)
\nonumber\\
&+&\nabla_{\vec{k}}~\epsilon(\vec{K})\cdot \nabla_{\vec{R}}~ 
G^{++}(\vec{R},T,\vec{k},\omega)~. 
\label{driving2}
\end{eqnarray}

To ensure gauge invariance of the kinetic equation we follow 
Ref.~\cite{Langreth66} and consider the generalized momentum
$\vec{K}=\vec{k}-(e/c)\vec{A}(\vec{R})$ as an independent variable instead
of the momentum $\vec{k}$. 
Using the two identities~\cite{Langreth66},
\begin{eqnarray}
\nabla_{\vec{R}}~a(\vec{K})&=&\nabla_{\vec{K}}~a(\vec{K})\cdot
\nabla_{\vec{R}}\vec{K}+
\nabla_{\vec{K}}~a(\vec{K})\times(\nabla_{\vec{R}}\times\vec{K})~, \nonumber\\
\vec{a}\cdot\vec{b}\times(\nabla_{\vec{R}}\times\vec{c})&=&
(\vec{a}\cdot\nabla_{\vec{R}}~\vec{c})\cdot\vec{b}-
(\vec{b}\cdot\nabla_{\vec{R}}~\vec{c})\cdot\vec{a}~,
\nonumber
\end{eqnarray}
Eq. (\ref{driving2}) becomes   
\begin{eqnarray}
D(\vec{R},T,\vec{K})=
\left(\partial_T + {e\over c} \nabla_{\vec{K}}~\epsilon(\vec{K})
\cdot (\vec{B} \times \nabla_{\vec{K}})
+\nabla_{\vec{K}}~\epsilon(\vec{K}) \cdot \nabla_{\vec{R}}~\right)
i \int_{-\infty}^{\infty}{{d\omega}\over{2\pi}}
G^{++}(\vec{R},T,\vec{K},\omega)~.
\label{driving3}
\end{eqnarray}

We assume weak interactions and replace the full Green functions by the
noninteracting Green functions (quasi-particle Ansatz),
\begin{eqnarray}
G^{pq}(\vec{R},T,\vec{K},\omega)=
G^{pq}_0(\vec{k}\rightarrow\vec{K},\omega)|_{N_0(\vec{k})\rightarrow N(\vec{R},T,\vec{K})}~, 
\label{replace}
\end{eqnarray}
where the noninteracting density matrix $N_0(\vec{k})$ is replaced by
the full density matrix $N(\vec{R},T,\vec{K})$. 
Performing the $\omega-$integrations in Eqs. (\ref{molecular}),
(\ref{collision}), and (\ref{driving3}) then yields 
the semiclassical kinetic equation for the electronic 
density matrix:
\begin{eqnarray}
& &\left(\partial_T + {e\over c}~\nabla_{\vec{K}}~\epsilon(\vec{K})\cdot
\vec{B} \times \nabla_{\vec{K}}
+\nabla_{\vec{K}}~\epsilon(\vec{K}) \cdot \nabla_{\vec{R}}\right)
N(\vec{R},T,\vec{K})=\nonumber\\
& &i[N(\vec{R},T,\vec{K}),\Delta^{++}(\vec{R},T,\vec{K})+
\tilde{\Sigma}^{++}(\vec{R},T,\vec{K},\epsilon(\vec{K}))]
\nonumber\\
& &+iN(\vec{R},T,\vec{K})\tilde{\Sigma}^{-+}(\vec{R},T,\vec{K},\epsilon(\vec{K}))
+i\tilde{\Sigma}^{+-}(\vec{R},T,\vec{K},\epsilon(\vec{K}))
[1-N(\vec{R},T,\vec{K})]
\label{Boltzmann0}~.
\end{eqnarray}                                 
To obtain a closed kinetic equation for the electronic density matrix, 
internal Green functions, which appear in the self-energies, have to be 
of course also eliminated according to Eq. (\ref{replace}). Details
concerning the calculation of self-energies are given in the next section.

For a homogeneous magnetic field, the electronic density matrix does not
explicitly depend on $\vec{R}$. The $\vec{R}$ dependence can be therefore 
neglected. For a quadratic dispersion, $\epsilon(\vec{K})=\vec{K}^2/2m^*$ ($\hbar=1$
in this subsection), the Lorentz term moreover becomes  
\begin{eqnarray}
(e/c)\nabla_{\vec{K}}~\epsilon(\vec{K})\times\vec{B}\cdot\nabla_{\vec{K}}
&=&(\vec{K}\times\vec{\Omega}_C)\cdot\nabla_{\vec{K}}
=-i\vec{\Omega}_C\cdot\vec{\cal L}~,
\end{eqnarray} 
where $\vec{\Omega}_C=e\vec{B}/m^*c$ is the cyclotron energy vector and
$\vec{\cal L}$ the angular momentum operator in $\vec{K}$-space, and we 
obtain the kinetic equation for the electronic density matrix in a more
familiar form:  
\begin{eqnarray}
\left( \partial_T - i \vec{\Omega}_C \cdot \vec{\cal L} \right)
N(T,\vec{K})&=&
i[N(T,\vec{K}),\Delta^{++}(T,\vec{K})+
\tilde{\Sigma}^{++}(T,\vec{K},\epsilon(\vec{K}))]
\nonumber\\
&+&iN(T,\vec{K})\tilde{\Sigma}^{-+}(T,\vec{K},\epsilon(\vec{K}))
\nonumber\\
&+&i\tilde{\Sigma}^{+-}(T,\vec{K},\epsilon(\vec{K}))
\large[1-N(T,\vec{K})\large]
\label{Boltzmann1}~.
\end{eqnarray}

This equation is the basis for the calculation of the spin relaxation time in
spatially homogeneous systems subject to a constant magnetic field.  
The first term on the rhs describes the coherent motion in a
molecular field modified by correlation effects.
If $\Delta^{++}$, $\tilde{\Sigma}^{++}$, and $N$ where scalar functions,
as in ordinary transport theory, this term would vanish.
The molecular field term is therefore a consequence of the
quantum mechanical treatment of the spin degree of freedom.  
To the singular part of the self-energy contribute the spin off-diagonal
terms in the Hamiltonian and the Hartree-Fock fields due to
electron-electron scattering. Dissipation and relaxation originate
from the regular part of the self-energy and give rise to the
second and third term on the rhs. They are at least
second order in the interaction. Formally,
they correspond to the scattering-out
and the scattering-in terms in a matrix-Boltzmann equation.
The matrix structure is of course a consequence of the full quantum
mechanical description of the spin. Only momentum scattering
is treated classically.

\subsection{Calculation of the self-energies}

\begin{figure}[t]
\hspace{0.0cm}\psfig{figure=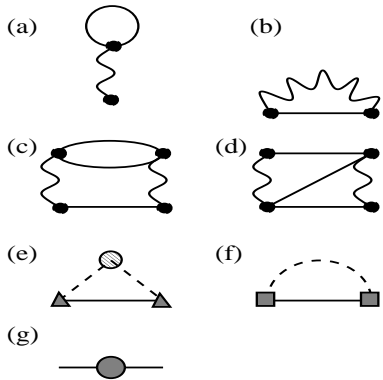,height=5.0cm,width=5.0cm,angle=0}
\caption[Diagram]
{Diagrammatic representation of self-energies in the Born approximation
for electron-electron (a)--(d), electron-impurity (e), and
electron-phonon scattering (f). Diagram (g) denotes the self-energy
due to spin off-diagonal Hamiltonian matrix elements.}
\label{Diagram}
\end{figure}

The semiclassical approach to furnish the self-energies in
the matrix-Boltzmann equation (\ref{Boltzmann1}),
valid for magnetic fields, which do not restructure the electron dispersion,
is to represent interaction processes in terms of diagrams,
calculate the diagrams using standard rules~\cite{Kinetic} to
obtain $\tilde{\Sigma}^{pq}(\vec{k},t,t')$ and $\Delta^{pq}(\vec{k},t)$,
perform the zeroth order gradient expansion, 
and then replace all internal Green functions according to 
Eq. (\ref{replace}). This heuristic strategy leads to self-energies, which
can be expressed in a 
manifestly gauge invariant form by writing the internal momentum 
integrations in terms of the generalized momentum $\vec{K}$.  
The formal structure of the self-energies is then 
the same as without magnetic field. Within the semiclassical approach,
the magnetic field gives therefore only rise to the Lorentz term. 

In Fig. \ref{Diagram} we depict the self-energies in the Born
approximation for electron-electron (a--d), electron-impurity (e),
and electron-phonon (f)
scattering. Diagram (g) corresponds to the self-energy due to the
spin off-diagonal term in the Hamiltonian (\ref{model}).

The Hartree-Fock diagrams (a) and (b) contribute to the instantaneous
self-energy $\Delta^{++}$. They are
second order in the spin polarization and therefore, for sufficiently
small spin polarizations, negligible. There are two second order
diagrams due to electron-electron scattering, the direct (c) and
the exchange (d) Born diagram. Anticipating that soft scattering
dominates, we neglect the exchange diagram (d).
The direct Born diagram contributes to $\tilde{\Sigma}^{++}$,
$\tilde{\Sigma}^{+-}$, and $\tilde{\Sigma}^{-+}$. It can be shown,
however, that $\tilde{\Sigma}^{++}$ is at least second order in the
spin polarization and therefore negligible in the limit of small
spin polarizations. The $\tilde{\Sigma}^{-+}$ and $\tilde{\Sigma}^{+-}$
components, contributing to the collision integral, are in contrast
linear in the spin polarization and cannot be neglected.
Diagrams (e) and (f), corresponding to the Born approximation for
electron-impurity and electron-phonon scattering, only contribute
to $\tilde{\Sigma}^{pq}$. As in the case of electron-electron
scattering, the $++$ component can be again neglected, if the spin
polarization
is small enough, whereas the $+-$ and $-+$ components contribute
in linear order in the spin polarization to the collision integral.
Diagram (g), corresponding to the spin off-diagonal part of the
Hamiltonian (\ref{model}), is linear in the spin polarization and
contributes to $\Delta^{++}$. Eventually it leads to a torque force
acting on the spin polarization. 

Anticipating small spin polarizations, we neglect $\tilde{\Sigma}^{++}$
and the Hartree-Fock contribution to ${\Delta}^{++}$. Writing
in the Born approximation furthermore
$I_B[N]=I_B^{ee}[N]+I_B^{ep}[N]+I_B^{ei}[N]$
for the second and third term on the rhs of Eq. (\ref{Boltzmann1}),
the semiclassical kinetic equation for the electronic density matrix 
reduces to
\begin{eqnarray}
\left( \hbar \partial_t
- i \hbar\vec{\Omega}_C \cdot \vec{\cal L} \right)N(\vec{k},t) =
{i\over 2}
\left[N(\vec{k}t),
\left(\hbar\vec{\Omega}_L+\hbar\vec{\Omega}_{IA}(\vec{k})
+\hbar\vec{\Omega}_{g}(\vec{k})\right) \cdot \vec{\sigma}\right]
+I_B[N]~,
\label{Boltzmann2}
\end{eqnarray}
where we relabeled the center time $T \rightarrow t$ and adjusted to the
notation of Eq. (\ref{DM}); $\hbar$ is explicitly included and the 
generalized momentum is now denoted by $\vec{k}$.

Equation (\ref{Boltzmann2}) is a matrix Boltzmann equation similar
to the semiconductor Bloch equations frequently used to describe
optically pumped semiconductors.\cite{HaugJauho} Thus, numerical
techniques used for the solution of the semiconductor Bloch
equations can be adopted to the numerical solution of Eq.
(\ref{Boltzmann2}). Calculations of this kind have been
successfully performed for various
situations.~\cite{Wu00a,Wu01,WuMetiu00,weng1,weng2,weng3}

To avoid a numerical solution, we focus on small spin
polarizations and linearize the Born collision integral with
respect to the spin polarization. It is important to note that the
equilibrium density matrix
$N_{eq}(\vec{k})=N(\vec{k},t\rightarrow\infty)$ is not diagonal in
the ``spin basis''. Expanding the equilibrium density matix in
terms of Pauli matrices yields
\begin{eqnarray}
N_{eq}(\vec{k})=f(\vec{k})+{1\over 2}\vec{\sigma}\cdot\vec{S}_{eq}(\vec{k})~,
\end{eqnarray}
where $f(\vec{k})=(1/2){\rm Tr}N_{eq}(\vec{k})=(f_+(\vec{k})+f_-(\vec{k}))/2$ is
half of the sum of the equilibrium distribution functions of the spin up
and spin down electrons and
$\vec{S}_{eq}(\vec{k})$ is the equilibrium spin polarization. Accordingly,
we also write for the density matrix at arbitrary times
\begin{eqnarray}
N(\vec{k},t)=f(\vec{k})+\delta f(\vec{k},t)+
{1\over 2}\vec{\sigma}\cdot[\vec{S}_{eq}(\vec{k})
+\delta\vec{S}(\vec{k},t)]~,
\label{Nexpansion}
\end{eqnarray}
with $\delta f(\vec{k},t)$ and $\delta\vec{S}(\vec{k},t)$ the
changes induced by optical pumping or by electrical injection. We
defined $N_{eq}(\vec{k})$ for $t\rightarrow \infty$, that is, it
contains the electrons created by the perturbation and both
$\delta \vec{S}(\vec{k},t)$ and $\delta f(\vec{k},t)$ have to
vanish for $t\rightarrow \infty$.

Inserting the expansion (\ref{Nexpansion}) into the Boltzmann
equation (\ref{Boltzmann2}) yields two kinetic equations, one for
the charge component $\delta f(\vec{k},t)$ and one for the spin
component $\delta\vec{S}(\vec{k},t)$. The collision terms couple
the two equations. If however only a small portion of the total
number of electrons initially contributed to the spin
polarization, i.e. if $\delta f(\vec{k},t) \ll f(\vec{k})$, the
coupling can be ignored and it suffices to focus on the equation
for $\delta \vec{S}(\vec{k},t)$ alone.

Since the total spin polarization $\vec{S}_{eq}(\vec{k})+\delta\vec{S}(\vec{k},t)$
is small, we linearize the Born collision integral with respect to both
$\vec{S}_{eq}(\vec{k})$ and $\delta\vec{S}(\vec{k},t)$. Thus,
\begin{eqnarray}
I_B[N]=I_B[N_{eq}]+I_B[f+\delta f,\delta\vec{S}]~.
\end{eqnarray}
If we now apply ${\rm Tr}\vec{\sigma}[...]$ on both sides of Eq. (\ref{Boltzmann2}),
use $I_B[N_{eq}]=0$ as well as $\vec{\Omega}_{IA}(\vec{k})\times\vec{S}_{eq}(\vec{k})=0$,
because, by construction, the equilibrium density matrix $N_{eq}(\vec{k})$ commutes
with $H_{IA}$, and ignore furthermore $\delta f$ in $I_B[f+\delta f,\delta\vec{S}]$, 
we get a closed kinetic equation for the non-equilibrium spin polarization 
\begin{eqnarray}
\left( \hbar\partial_t - i \hbar\vec{\Omega}_C \cdot \vec{\cal L}
\right) \delta\vec{S}(\vec{k},t)=
\left(\hbar\vec{\Omega}_L+
\hbar\vec{\Omega}_{IA}(\vec{k})+\hbar\vec{\Omega}_{g}(\vec{k})\right)
\times
\delta\vec{S}(\vec{k},t)+J_B[f,\delta\vec{S}]
\label{KEQS1}
\end{eqnarray}
with $J_B[f,\delta\vec{S}]={\rm Tr}\vec{\sigma}I_B[f,\delta\vec{S}]$.

The collision integral can be further simplified if we split the spin-flip
matrix into a leading spin conserving diagonal part and a small off-diagonal
part which describes spin-flip scattering. Since nonparabolicities are small,
the diagonal part is approximately equal to the unit matrix and we get
$I(\vec{k},\vec{k'})\simeq 1+\delta I(\vec{k},\vec{k'})$, with
$\delta I(\vec{k},\vec{k'})\ll 1$. Expanding the collision integrals
up to second order in $\delta I(\vec{k},\vec{k'})$ gives
\begin{eqnarray}
J_B[f,\delta\vec{S}]=J_B[f,\delta\vec{S}]^{(0)}+
J_B[f,\delta\vec{S}]^{(1)}+J_B[f,\delta\vec{S}]^{(2)}~.
\label{KEQS2}
\end{eqnarray}
The first order term $J_B^{(1)}[f,\delta\vec{S}]$ potentially
mixes EY, DP, and VG spin relaxation channels, but for
semiconductors with high symmetry it does not contribute to the
spin relaxation rates.

The kinetic equation for the excess spin polarization becomes therefore
\begin{eqnarray}
\left( \hbar\partial_t - i \hbar \vec{\Omega}_C \cdot \vec{\cal L} \right)
\delta\vec{S}(\vec{k},t)&=&
\hbar\vec{\Omega}_L\times\delta\vec{S}(\vec{k},t)
+\left(\hbar\vec{\Omega}_{IA}(\vec{k})+\hbar\vec{\Omega}_g(\vec{k})\right)
\times
\delta\vec{S}(\vec{k},t)
\nonumber\\
&+&J_B^{(0)}[f,\delta\vec{S}]+
J_B^{(2)}[f,\delta\vec{S}]~.
\label{Boltzmann3}
\end{eqnarray}
This equation contains motional-narrowing (DP and VG) and spin-flip (EY)
spin relaxation processes on an
equal footing. The Elliott-Yafet process is simply encoded in
$J_B^{(2)}[f,\delta\vec{S}]$  whereas the motional-narrowing processes
result from the combined action of the
torque forces given by the second term on the rhs and the spin conserving
scattering processes comprising $J_B^{(0)}[f,\delta\vec{S}]$.

Independent of the scattering process, the structure of the collision
integrals in  Eq. (\ref{Boltzmann3}) is ($\nu$ = ei, ee, and ep)
\begin{eqnarray}
J_\nu^{(0)}[f,\delta\vec{S}]&=&\sum_{ \vec{q} } [
W^\nu(\vec{k}+\vec{q};\vec{q})~\delta\vec{S}(\vec{k}+\vec{q},t)-
W^\nu(\vec{k};\vec{q})~\delta\vec{S}(\vec{k},t) ]~,
\label{CI0}\\
J_\nu^{(2)}[f,\delta\vec{S}]&=&2\sum_{\vec{q}}W^\nu(\vec{k}+\vec{q};\vec{q})
~\vec{g}(\vec{k},\vec{k}+\vec{q})\times
[\vec{g}(\vec{k},\vec{k}+\vec{q})\times\delta\vec{S}(\vec{k}+\vec{q},t)]~,
\label{CI2}
\end{eqnarray}
where, for concise notation, we introduced a spin-flip vector
\begin{eqnarray}
\vec{g}(\vec{k},\vec{k}^{'})=
\left(\begin{array}{c}
{\rm Im}I_{+-}(\vec{k},\vec{k}^{'})\\
{\rm Re}I_{+-}(\vec{k},\vec{k}^{'})\\
0
\end{array}\right)~,
\end{eqnarray}
with $I_{+-}(\vec{k},\vec{k}^{'})$ the off-diagonal element of
the overlap matrix (\ref{overlap}). This is a result of the
Born approximation and the linearization with respect to the spin
polarization. In general, the structure of the collision integrals depends
on the scattering process. Here, however, the scattering process
enters only through $W^\nu(\vec{k};\vec{q})$, the probabilities for
a transition between momentum state $\vec{k}-\vec{q}$ and $\vec{k}$.
For electron-ionized-impurity scattering, for instance,
\begin{eqnarray}
W^{ei}(\vec{k};\vec{q})=2 \pi N_i |U(q)|^{2}
\delta(\varepsilon(\vec{k}-\vec{q})-\varepsilon(\vec{k}))~,
\label{Wei}
\end{eqnarray}
while for electron-electron scattering,
\begin{eqnarray}
W^{ee}(\vec{k};\vec{q})&=&4\pi |V(q)|^{2}\sum_{\vec{k}^{'}}
      \bigg\{[1-f(\vec{k}-\vec{q})-f(\vec{k}^{'}+\vec{q})]f(\vec{k}^{'})
\nonumber\\
      &+&f(\vec{k}-\vec{q})f(\vec{k}^{'}+\vec{q})\bigg\}
\delta(\varepsilon(\vec{k})+\varepsilon(\vec{k}^{'})
      -\varepsilon(\vec{k}-\vec{q})-\varepsilon(\vec{k}^{'}+\vec{q}))~,
\label{Wee}
\end{eqnarray}
with $U(q)$ and $V(q)$ statically screened Coulomb potentials.
Similar expressions hold for electron-phonon scattering. For electron
impurity scattering, which is elastic, $W^{ei}(\vec{k}+\vec{q};\vec{q})=
W^{ei}(\vec{k};\vec{q})$; moreover $W^{ei}(\vec{k};\vec{q})$ is
independent of the equilibrium distribution of the spin-up and the
spin-down electrons. In
general, however, the transition probabilities
depend on the equilibrium distribution of the electrons, and, in the case
of electron-phonon scattering, also on the equilibrium distribution of
the phonons.

\subsection{Diffusion approximation}

The simple form of the collision integrals (\ref{CI0}) and
(\ref{CI2}) suggests to conceptualize the dynamics of the
non-equilibrium spin polarization in terms of spin-polarized
``test'' electrons, scattering off an equilibrated 
bath of ``field'' particles (impurities, electrons, and phonons).
Usually this picture can be only applied
to electron-impurity and electron-phonon scattering, where the
scattering partners belong to different species, and not to
electron-electron scattering, where the scattering partners belong
to the same species. It is only within the linearized spin dynamics, which
essentially treats the electrons comprising the non-equilibrium spin 
polarization as a
separate species, that the ``test-field-particle concept'' can be 
applied to electron-electron scattering as well.
We now take full advantage of the simplicity of the collision integrals
and expand the collision integrals with respect to the momentum
transfer $\vec{q}$. As a result the integro-differential equation
(\ref{Boltzmann3}) becomes a differential equation.

The on-shell spin-conserving process due to elastic
electron-impurity scattering yields
\begin{eqnarray}
J_B^{\rm (0),on}[f,\delta\vec{S}]=\sum_n\sum_{i_1,...,i_n}
C^{\rm ei}_{i_1,...,i_n}(\vec{k})
{\partial^n\over{\partial {k_{i_1}}...\partial {k_{i_n}}}}
\delta\vec{S}(\vec{k},t)~,
\label{diff1}
\end{eqnarray}
whereas the inelastic spin-conserving processes due to
electron-electron or electron-phonon scattering give rise to
an off-shell contribution 
\begin{eqnarray}
J_B^{\rm (0),off}[f,\delta\vec{S}]=
\sum_{\nu={\rm ee,ep}}
\sum_n\sum_{i_1,...,i_n}
{\partial^n\over{\partial_{k_{i_1}}...\partial {k_{i_n}}}}
C^{\rm \nu}_{i_1,...,i_n}(\vec{k})\delta\vec{S}(\vec{k},t)~,
\label{diff2}
\end{eqnarray}
where, in both cases, the moments are defined by ($\nu={\rm ei,ee,ep}$)
\begin{eqnarray}
C^\nu_{i_1,...,i_n}(\vec{k})={1\over n!}\sum_{\vec{q}}
q_{i_1}...q_{i_n}W^\nu(\vec{k};\vec{q})~.
\label{diff3}
\end{eqnarray}
The transition probability $W^\nu(\vec{k};\vec{q})$ depends
on the precise modeling of the elementary scattering process and
also on the dimensionality of the system. In the Appendix we give
explicit expressions for electron-electron
and electron-impurity scattering in a quantum well. Note, for
inelastic scattering the differential operators act on the
moments $C^\nu_{i_1,...,i_n}(\vec{k})$ whereas for elastic
scattering the moments are in front of the differential operators.


Equations (\ref{diff1}) and (\ref{diff2}) involve partial differential
operators of arbitrary order. To obtain tractable equations,
the expansion is in many cases truncated after the second order
term (diffusion approximation). As a result, scattering processes
with small momentum transfer are treated exactly whereas
scattering processes with large momentum transfer are
treated approximately. Because the transition
probability for the (unscreened) Coulomb potential diverges for
small momentum transfer, soft Coulomb scattering events
dominate, and the diffusion approximation is expected to describe
Coulomb scattering reasonably well. A similar
reasoning applies also to electron-LO-phonon scattering. The
singular behavior of the collision integrals is stronger in
three than in two dimensions. The diffusion approximation is 
therefore somewhat better justified for bulk than for quantum well 
situations.~\cite{Glazov03a,Glazov02,Glazov03b,diffapprox} 
Nevertheless, our numerical results suggest 
that even for a quantum well the diffusion approximation gives 
reasonable quantitative results for the spin relaxation time.

Keeping therefore only the second order terms, we write 
\begin{eqnarray}
J_B^{(0)}[f,\delta\vec{S}]&=&J_B^{\rm (0),on}[f,\delta\vec{S}]+
J_B^{\rm (0),off}[f,\delta\vec{S}]\nonumber\\
&=&\left[\sum_i A^{\rm ei}(\vec{k}){\partial\over{\partial {k_i}}}
+\sum_{ij}B_{ij}^{\rm ei}(\vec{k}){\partial^2\over{\partial {k_i}\partial {k_j}}}
\right]\delta\vec{S}(\vec{k},t)\nonumber\\
&+&\sum_{\nu={\rm ee,ep}}\left[
\sum_i{\partial\over{\partial {k_i}}}A^\nu _i(\vec{k})
+\sum_{ij}{\partial^2\over{\partial {k_i}\partial {k_j}}}
B_{ij}^\nu(\vec{k})\right]\delta\vec{S}(\vec{k},t)\nonumber\\
&=&{\cal D}(\vec{k})\delta\vec{S}(\vec{k},t)~,
\label{diffapp}
\end{eqnarray}
where the first two terms on the rhs come from elastic
scattering processes and the last two terms encode inelastic
scattering events. In Eq. (\ref{diffapp}) we introduced for the
first and second moments ($\nu={\rm ei,ee,ep}$),
\begin{eqnarray}
A^\nu_{i}(\vec{k})&=&\sum_{\vec{q}}
q_i W^\nu(\vec{k};\vec{q})~,
\label{Ai}\\
B^\nu_{ij}(\vec{k})&=&{1\over 2}\sum_{\vec{q}}
q_i q_j W^\nu(\vec{k};\vec{q})~,
\label{Bij}
\end{eqnarray}
which have the meaning of $\vec{k}$-dependent dynamical friction
and diffusion coefficients, respectively. Within the diffusion
approximation the spin-conserving (Born) collision integrals are
therefore represented by a Fokker-Planck differential operator
(\ref{diffapp}).
Each scattering process
gives rise to a particular Fokker-Planck operator, with 
particular dynamical friction
and diffusion coefficients.

In the same spirit, expanding the spin-flip collision integral
$J_B^{(2)}[f,\delta\vec{S}]$ up to
second order in the momentum transfer $\vec{q}$, and using
$\vec{g}(\vec{k},\vec{k})=0$, gives
\begin{eqnarray}
J_B^{(2)}[f,\delta\vec{S}]=-{\bf
R}(\vec{k})\delta\vec{S}(\vec{k}t)~,
\label{JB2}
\end{eqnarray}
with a spin-flip tensor
\begin{eqnarray}
{\bf R}(\vec{k})=4\sum_{\nu={\rm ei,ee,ep}}\sum_{ij}B^\nu_{ij}(\vec{k}){\bf
G}_{ij}(\vec{k})~,
\label{GammaEY0}
\end{eqnarray}
given in terms of the total diffusion coefficient and a
tensor ${\bf G}_{ij}(\vec{k})$ which describes the rate of change
of the spin-flip vector $\vec{g}(\vec{k},\vec{k}')$:
\begin{eqnarray}
{\bf G}_{ij}(\vec{k})=\left( \begin{array}{ccc}
O_{ij}^{\rm yy} & -O_{ij}^{\rm xy} & 0 \\
-O_{ij}^{\rm yx} & O_{ij}^{\rm xx} & 0  \\
0 & 0 & O_{ij}^{\rm xx} + O_{ij}^{\rm yy}
\end{array} \right) ,
\end{eqnarray}
with
\begin{eqnarray}
O_{ij}^{\nu\mu}=\left[{\partial\over{\partial k_i}}
g^\nu_{\vec{k}\vec{k}^{'}}\right]_{\vec{k}^{'}=\vec{k}} \left[{\partial
\over{\partial k_j}} g^\mu_{\vec{k}\vec{k}^{'}}\right]_{\vec{k}^{'}=\vec{k}} .
\end{eqnarray}

\begin{figure}[t]
\hspace{0.0cm}\psfig{figure=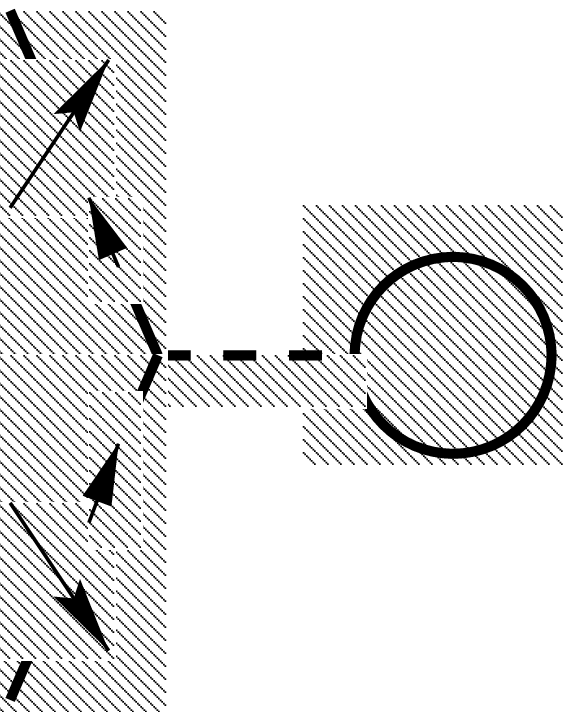,height=3.0cm,width=2.5cm,angle=0}
\caption
{Graphical illustration of the collision terms within the
diffusion approximation: A ``test'' spin polarization
scatters off a generalized  bath of equilibrated ``field'' particles
(electrons, impurities, and/or phonons). Whereas
spin-flip scattering is elastic (on-shell) within our
approximation, spin-conserving scattering
can be on- or off-shell, i.e. the ``test'' spin polarization
can lose/gain energy, because the ``internal degrees"
of the bath can absorb/emit energy.
}
\label{Cartoon}
\end{figure}
In Fig \ref{Cartoon} we illustrate the physical content of the
diffusion approximation encoded in
Eqs. (\ref{diffapp}) and
(\ref{JB2}): the small ``test'' spin polarization
$\delta\vec{S}(\vec{k},t)$ scatters off equilibrated ``field''
particles, which, depending on the scattering process, are either
electrons, phonons, or impurities. Spin conserving scattering
can be elastic and inelastic, because the ``field'' particles
can absorb/emit energy, the bath has ``internal degrees''.
Spin non-conserving scattering, on the other hand, turns out to
be elastic within the diffusion approximation.

We now introduce scaled atomic units and measure energy in units of a
scaled atomic Rydberg $\tilde{R}_0=R_0/s$ and
length in units of a scaled atomic Bohr radius $\tilde{a}_0=\sqrt{s}a_0$, with
$\tilde{R}_0\tilde{a}_0^2=\hbar^2/2m_0$ and $e^2=2\sqrt{s}\tilde{R}_0\tilde{a}_0$,
where $m_0$ is the bare electron
mass and $s$ is a scale factor chosen to yield $\tilde{R}_0=1~ {\rm meV}$.
Symmetry-adapted coordinates are then a radial coordinate $\varepsilon=k^2$
and a generalized angle variable $\omega$, which, for bulk semiconductors
comprises two angles, the polar angle $\theta$ and the azimuth angle $\phi$,
and for quantum wells is simply the polar angle $\phi$. Before we
express the Fokker-Planck equation in these symmetry adopted coordinates,
we recall that the experimentally measured quantity is the macroscopic
($\vec{k}$-averaged) spin polarization. Normalizing the
macroscopic spin polarization to $N_s$, the (small) number of initially
spin-polarized electrons and writing the $\vec{k}$-integral in
symmetry-adapted coordinates, we define a ``macroscopic''
spin polarization (per spin polarized electron) ,
\begin{eqnarray}
\delta \vec{S}(t)&=&
{1\over N_s}\sum_{\vec{k}}\delta\vec{S}(\vec{k},t)
\nonumber\\
&=&{1\over{(2\pi)^d n_s}}
\int_0^\infty d\varepsilon \int d\omega~J(\varepsilon)
\delta\vec{S}(\varepsilon,\omega,t)~,
\label{totaltest}
\end{eqnarray}
with $d$ the dimension, $n_s=N_s/L^d$ the density of initially spin-polarized
electrons and
$J(\varepsilon)$ the energy dependent part of the Jacobian, which arises
from the transformation to the symmetry-adapted coordinates.
Note that due to the normalization,
$\delta \vec{S}(0)$ is a unit vector in the direction of the initial
spin polarization. For bulk,
$d=3$, $d\omega=d\phi d\theta \sin{\theta}$, and
$J(\varepsilon)=\sqrt{\varepsilon}/2$, while for quantum wells $d=2$,
$d\omega=d\phi$, and $J(\varepsilon)=1/2$. Instead of
setting up the Fokker-Planck equation for $\delta\vec{S}(\varepsilon,\omega,t)$
it is more convenient to directly construct the Fokker-Planck equation for
\begin{eqnarray}
\delta\vec{S}'(\varepsilon,\omega,t)
={{J(\varepsilon)}\over (2\pi)^d n_s}\delta\vec{S}(\varepsilon,\omega,t)~.
\end{eqnarray}

The differential operator describing spin conserving scattering processes
in the Fokker-Planck equation for $\delta\vec{S}'$ reads in symmetry-adapted
coordinates
\begin{eqnarray}
{\cal D}(\vec{k})&=&-{\partial\over{\partial\varepsilon}}
{\varepsilon\over{\tau_f(\varepsilon)}}+{\partial^2\over{\partial\varepsilon^2}}
{\varepsilon^2\over{\tau_d(\varepsilon)}}-{1\over{4\tau_\perp(\varepsilon)}}
{\cal L}^2(\omega)\nonumber\\
&=&{\cal D}(\varepsilon)-{1\over{4\tau_\perp(\varepsilon)}}
{\cal L}^2(\omega)~,
\label{diffop}
\end{eqnarray}
where the operator ${\cal L}^2(\omega)$ denotes the total
angular momentum operator in momentum space.
To obtain this generic form for both bulk and quantum wells it is essential
to include $J(\varepsilon)$ into the definition of the spin polarization.
The off-shell term ${\cal D}(\varepsilon)$ originates from inelastic scattering
events, e.g., due to electron-electron or electron-phonon scattering.
The relaxation rates $1/\tau_f(\varepsilon)$ and $1/\tau_d(\varepsilon)$
denote the rate with which the ``test'' spin polarization loses energy and
the rate with which the ``test'' spin polarization diffuses in energy space,
respectively. The on-shell term, describing randomization of the angle
variable, is given by the last term on the rhs of Eq. (\ref{diffop}).
It is proportional to the total on-shell relaxation rate
$1/\tau_\perp(\varepsilon)$ due to both elastic and inelastic
scattering processes.

The rates characterizing the differential operator ${\cal D}(\vec{k})$
are obtained from a direct calculation
of the coefficients $A^\nu_i(\vec{k})$ and $B^\nu_{ij}(\vec{k})$ and
casting the resulting differential operator ${\cal D}(\vec{k})$ in the specific
form given in Eq. (\ref{diffop}). An explicit calculation of the
symmetry-adapted form of the relaxation tensor ${\bf R}(\vec{k})$, which
is defined in terms of the total diffusion coefficient
$B_{ij}(\vec{k})$, due to both elastic and inelastic scattering
processes, shows moreover
that it can be expressed in terms of the same scattering rates.
Thus, the three scattering rates $1/\tau_f(\varepsilon)$,
$1/\tau_d(\varepsilon)$,
and $1/\tau_\perp(\varepsilon)$ completely specify the two collision
integrals $J_B^{(0)}[f,\delta\vec{S}]$ and $J_B^{(2)}[f,\delta\vec{S}]$.
In the Appendix we give explicit expressions for the relaxation
rates due to electron-electron and electron-impurity scattering in a
quantum well.

The dimensionless, symmetry-adapted Fokker-Planck equation for
$\delta\vec{S}'$, which is the basis for the calculation of the
spin relaxation rates presented in the next subsection, can be
therefore written as
\begin{eqnarray}
{\partial\over{\partial t}}\delta\vec{S}'(\varepsilon,\omega,t)&=&
\large[{\cal D}(\varepsilon)
-{1\over{4\tau_\perp(\varepsilon)}}{\cal L}^2(\omega)
+i\vec{\Omega}_C\cdot\vec{{\cal L}}(\omega)\large]
\delta\vec{S}'(\varepsilon,\omega,t)
+\vec{\Omega}_L\times\delta\vec{S}'(\varepsilon,\omega,t)
\nonumber\\
&+&\large[\vec{\Omega}_{IA}(\varepsilon,\omega)
+\vec{\Omega}_{g}(\varepsilon,\omega)\large]\times\delta\vec{S}'(\varepsilon,\omega,t)
-{\bf R}(\varepsilon,\omega)~\delta\vec{S}'(\varepsilon,\omega,t)~.
\label{FPE}
\end{eqnarray}
It contains spin relaxation due
to motional-narrowing (DP and VG processes) and spin-flip scattering
(EY process). The former arises from
the combined action of the off- and on-shell spin-conserving scattering
events encoded in the differential operators ${\cal D}(\varepsilon)$ and
$(1/\tau_\perp(\omega)){\cal L}^2(\omega)$, respectively, and the
torque forces due to
$\vec{\Omega}_{IA}(\varepsilon,\omega)$ and $\vec{\Omega}_g(\varepsilon,\omega)$,
while the latter originates from the spin-flip tensor
${\bf R}(\varepsilon,\omega)$.
Note that if more than one scattering process is considered,
the relaxation rates characterizing the differential
operators and the spin-flip tensor are total relaxation rates, due to
whatever elastic and inelastic scattering processes are included in the
model. The orbital motion of the electrons, which leads to a quenching of the
motional-narrowing spin relaxation processes, is given by the term
$i\vec{\Omega}_C\cdot\vec{\cal L}(\omega)$ on the rhs of Eq. (\ref{FPE}).
In the next subsection
we develop a scheme which separates the fast spin conserving
scattering processes from the slow spin decay causing processes and
directly yields the time evolution of the macroscopic spin polarization.

\subsection{Multiple time scale analysis}

The Fokker-Planck equation (\ref{FPE})
determines the time evolution of the non-equilibrium
spin polarization on the fast, spin-conserving time scale, where
randomization of the angle variables (direction of the momentum) and
energy relaxation and diffusion
occurs, and on the long time scale, where spin non-conserving processes lead to
the decay of the spin polarization. The two time scales are well separated.
The fast, spin conserving stage, whose scale is given by the first term on
the rhs of Eq. (\ref{FPE}) and therefore by the off- and on-shell
relaxation times (as well as the time it takes to complete
one cyclotron orbit), terminates in a quasi-stationary state, which then
evolves on the time scales set by the Elliott-Yafet term, the
torque forces due to the spin off-diagonal Hamiltonian matrix elements,
and the external magnetic field (Larmor precession).
Experimentally relevant is usually the time evolution on the long time scale.
In this subsection we employ therefore a multiple time scale approach to
extract from
the Fokker-Planck equation (\ref{FPE}) a Bloch equation, which controls the
time evolution of the macroscopic ($\vec{k}$-averaged)
spin polarization on the long time scale.
To simplify the notation, we suppress the prime. It is
understood that the spin polarization $\delta\vec{S}(\varepsilon,\omega,t)$
contains the factor $J(\varepsilon)/(2\pi)^d n_s$.

As a preparatory step we first consider that part of the
Fokker-Planck equation (\ref{FPE}), which is spin-conserving:
\begin{eqnarray}
{\partial\over{\partial t}}\delta\vec{S}(\varepsilon,\omega,t)=
[{\cal D}(\varepsilon)
-{1\over{4\tau_\perp(\varepsilon)}}{\cal L}^2(\omega)
+i\vec{\Omega}_C\cdot\vec{{\cal L}}(\omega)]\delta\vec{S}(\varepsilon,\omega,t)~.
\label{FPEcon}
\end{eqnarray}
To find the stationary solution of Eq. (\ref{FPEcon}) we set the lhs
to zero, write
\begin{eqnarray}
\delta \vec{S}_{st}(\varepsilon,\omega)=p(\varepsilon)
\vec{\Psi}_{st}(\varepsilon,\omega)~,
\label{deltaSinit}
\end{eqnarray}
with $p(\varepsilon)$ defined by
\begin{eqnarray}
{\cal D}(\varepsilon)p(\varepsilon)=0~,
\label{pdiff}
\end{eqnarray}
and obtain for the auxiliary vector $\vec{\Psi}_{st}(\varepsilon,\omega)$
the differential equation
\begin{eqnarray}
[{\cal D}^*(\varepsilon)
-{1\over{4\tau_\perp(\varepsilon)}}{\cal L}^2(\omega)
+i\vec{\Omega}_C\cdot\vec{{\cal L}}(\omega)]\vec{\Psi}_{st}(\varepsilon,\omega)=0~,
\label{FPEstat}
\end{eqnarray}
with
\begin{eqnarray}
{\cal D}^*(\varepsilon)=
{\varepsilon\over{\tau_f(\varepsilon)}}{\partial\over{\partial\varepsilon}}
+{\varepsilon^2\over{\tau_d(\varepsilon)}}{\partial^2\over{\partial\varepsilon^2}}~,
\end{eqnarray}
the adjoint operator to ${\cal D}(\varepsilon)$.
The simplest solution of (\ref{FPEstat}) is a constant vector
\begin{eqnarray}
\vec{\Psi}_{st}(\varepsilon,\omega)=\vec{e}~,
\end{eqnarray}
giving rise to a stationary solution
\begin{eqnarray}
\delta\vec{S}_{st}(\varepsilon,\omega)=p(\varepsilon)\vec{e}~.
\label{Sst}
\end{eqnarray}
The particular form of $p(\varepsilon)$ does not matter at this
point.

We now turn to the full Fokker-Planck equation (\ref{FPE}).
The spin-conserving stage of the time evolution, described by
the first term on the rhs, occurs on a very
fast time scale and is usually experimentally not
resolved. Hence, it is not necessary to explicitly keep track of it.
Instead, it is sufficient to use the
final state of the fast, spin-conserving time evolution, i.e. the
stationary solution of the spin conserving part of the Fokker-Planck
equation (viz: Eqs. (\ref{FPEcon})--(\ref{Sst})), as an
initial state for the time evolution on the slow time scale, where spin
decay occurs. We write the initial condition therefore as
\begin{eqnarray}
\delta\vec{S}_{st}(\varepsilon,\omega,0)=p(\varepsilon)\vec{e}~,
\end{eqnarray}
where $\vec{e}$ is now the direction of the initial spin polarization.
This initial condition is general enough,
because neither electrical nor optical spin
injection produce anisotropic initial spin
polarizations. Accordingly, we write for arbitrary times
\begin{eqnarray}
\delta \vec{S}(\varepsilon,\omega,t)=p(\varepsilon)\vec{\Psi}(\varepsilon,\omega,t)~,
\label{deltaS}
\end{eqnarray}
where the time-dependent auxiliary vector $\vec{\Psi}(\varepsilon,\omega,t)$
satisfies now the time-dependent equation
\begin{eqnarray}
{\partial\over{\partial t}}\vec{\Psi}(\varepsilon,\omega,t)&=&
[{\cal D}^*(\varepsilon)-{1\over{4\tau_\perp(\varepsilon)}}{\cal L}^2(\omega)
+i\vec{\Omega}_C\cdot\vec{{\cal L}}(\omega)]
\vec{\Psi}(\varepsilon,\omega,t)
\nonumber\\
&+&\left(\vec{\Omega}_L+\vec{\Omega}_{IA}(\varepsilon,\omega)
+\vec{\Omega}_g(\varepsilon,\omega)\right) \times
\vec{\Psi}(\varepsilon,\omega,t)
-{\bf R}(\varepsilon,\omega)\vec{\Psi}(\varepsilon,\omega,t)~,
\label{FPad}
\end{eqnarray}
with an initial condition $\vec{\Psi}(\varepsilon,\omega,0)=\vec{e}$.
Note that Eq. (\ref{FPad}) is the adjoint Fokker-Planck equation.
The function $p(\varepsilon)$ satisfies the homogeneous
differential equation (\ref{pdiff}) and is therefore defined only up
to a normalization constant. From the initial condition for the 
macroscopic spin polarization, $\delta\vec{S}(0)=\vec{e}$,
we conclude that $p(\varepsilon)$ has to be normalized according to 
$\int d\varepsilon \int d\omega p(\varepsilon)=1$
(recall that we redefined $\delta\vec{S}$
such that it contains the factor $J(\varepsilon)/(2\pi)^d n_s$).
Thus, $d\varepsilon\int d\omega p(\varepsilon)$ can be
interpreted as the probability density for finding a spin-polarized
``test'' electron in the energy interval 
$[\varepsilon,\varepsilon+d\varepsilon]$. 

To proceed further we scale each term in Eq. (\ref{FPad}) to its
typical value. In the case of degenerate electrons the typical values would be
the ones at the Fermi energy, whereas for non-degenerate electrons the
typical values could be the ones at the average thermal energy. Denoting
typical values by a caret, we introduce scaled quantities $t'=t/\hat{t}$,
$\tau_f'=\tau_f/\hat{\tau}_f$, $\tau_d'=\tau_d/\hat{\tau}_d$,
$\tau_\perp'=\tau_\perp/\hat{\tau}_\perp$,
$\vec{\Omega}'_{IA}=\vec{\Omega}_{IA}\hat{\tau}_{IA}$,
$\vec{\Omega}'_g=\vec{\Omega}_g\hat{\tau}_g$,
$\vec{\Omega}_L'=\vec{\Omega}_L\hat{\tau}_L$,
${\bf R}'={\bf R}\hat{\tau}_R$,
and $\vec{\Omega}_C'=\vec{\Omega}_C\hat{\tau}_C$.
The rescaled equation
for $\vec{\Psi}'(\varepsilon,\omega,t')$ becomes (suppressing the
arguments of the various functions)
\begin{eqnarray}
{\partial\over{\partial t'}}\vec{\Psi}'&=&
[{\hat{t}\over{\hat{\tau}_f}}
{\varepsilon\over{\tau_f'}}{\partial\over{\partial\varepsilon}}
+{\hat{t}\over{\hat{\tau}_d}}
{\varepsilon^2\over{\tau_d'}}{\partial^2\over{\partial\varepsilon^2}}
-{\hat{t}\over{\hat{\tau}_\perp}}
{1\over{4\tau_\perp'}}{\cal L}^2
+{\hat{t}\over{\hat{\tau}_C}}
i\vec{\Omega}_C'\cdot\vec{{\cal L}}]\vec{\Psi}'\nonumber\\
&+&{\hat{t}\over{\hat{\tau}_L}}\vec{\Omega}_L'\times\vec{\Psi}'
+{\hat{t}\over{\hat{\tau}_{IA}}}
\vec{\Omega}'_{IA}\times\vec{\Psi}'
+{\hat{t}\over{\hat{\tau}_g}}
\vec{\Omega}'_g\times\vec{\Psi}'
-{\hat{t}\over{\hat{\tau}_R}}
{\bf R}'\vec{\Psi}'~.
\label{hhhh}
\end{eqnarray}
We identify three time scales.
A fast time scale given by the spin conserving relaxation times
$\hat{\tau_i}$ ($i = f, d, \perp$) and the time it takes to complete
a cyclotron orbit $\hat{\tau}_C$, an intermediate time scale given by
the time it takes to complete a precession around the intrinsic magnetic
fields (due to the spin off-diagonal Hamiltonian matrix elements)
$\hat{\tau}_{IA}$ and $\hat{\tau}_g$, and a long time scale,
on which Larmor precession and spin-flip scattering occur,
$\hat{\tau}_L$ and $\hat{\tau}_R$, respectively.
For representative experimental set-ups, the typical time scale 
$\hat{t}$, on which the spin polarization has to be tracked 
(``observation time''), and the three typical intrinsic time scales
obey the following ordering:
${\hat{t}\over{\hat{\tau}_f}}, {\hat{t}\over{\hat{\tau}_d}},
{\hat{t}\over{\hat{\tau}_\perp}}, {\hat{t}\over{\hat{\tau}_C}}=O(\eta^{-1})$,
${\hat{t}\over{\hat{\tau}_{IA}}},
{\hat{t}\over{\hat{\tau}_g}}=O(\eta^{0})$, and
${\hat{t}\over{\hat{\tau}_{L}}},{\hat{t}\over{\hat{\tau}_{R}}}=O(\eta^{1})$
where we introduced a small parameter $\eta$. Accordingly, we classify
each term in Eq. (\ref{hhhh}) by the smallness
parameter $\eta$. Suppressing the primes, Eq. (\ref{hhhh})
is rewritten as
\begin{eqnarray}
{\partial\over{\partial t}}\vec{\Psi}=
{1\over\eta}[{\cal D}^*-{1\over{4\tau_\perp}}{\cal L}^2
+i\vec{\Omega}_C\cdot\vec{{\cal L}}]
\vec{\Psi}
+\large(\vec{\Omega}_{IA}
+\vec{\Omega}_g\large)\times\vec{\Psi}
+\eta~\vec{\Omega}_L\times\vec{\Psi}
-\eta~{\bf R}\vec{\Psi}~.
\label{FPad2}
\end{eqnarray}

Equation (\ref{FPad2}) is in a form where fast and slow processes can be
clearly identified. The fast spin-conserving terms and the orbital motion
enter in order $\eta^{-1}$, the precession around the internal magnetic fields
enters in order $\eta^0$, whereas the Larmor precession and the spin-flip
scattering terms appear in order $\eta^1$. Naturally, taking as much
advantage as possible of the existence of the small parameter $\eta$,
the first thought is to expand all quantities with respect to $\eta$ and
apply perturbation theory.
The structure of Eq. (\ref{FPad2}) indicates however that regular
perturbation theory will lead to non-uniformity in the long-time regime,
i.e. precisely in that regime,  which we are interested in. To obtain the
correct long-time behavior of the solution of Eq. (\ref{FPad2})
a multiple time scale approach is required.

In the spirit of multiple time scale perturbation theory~\cite{Nayfeh},
we consider therefore $\vec{\Psi}$
as a function of three time variables $t_n = \eta^n t$, $n = -1, 0, 1$,
which are assumed to be independent, and substitute a second order expansion
of the form
\begin{eqnarray}
\vec{\Psi}(\varepsilon,\omega,t)&=&\vec{\Psi}^{(0)}(\varepsilon,\omega,t_{-1},t_0,t_1)
+\eta \vec{\Psi}^{(1)}(\varepsilon,\omega,t_{-1},t_0,t_1)
+\eta^2 \vec{\Psi}^{(2)}(\varepsilon,\omega,t_{-1},t_0,t_1)
\label{PsiExp}
\end{eqnarray}
into Eq. (\ref{FPad2}), where the time derivative is extended to
$\partial_t=\eta^{-1}\partial_{t_{-1}} + \partial_{t_0} + \eta \partial_{t_1}$.
Equating coefficients of like powers of $\eta$ yields an hierarchy of equations
for the functions $\vec{\Psi}^{(n)}$. Up to $O(\eta)$ they read
\begin{eqnarray}
{\partial\over{\partial t_{-1}}}\vec{\Psi}^{(0)}&=&[{\cal D}^*-
{1\over{4\tau_\perp}}{\cal L}^2
+i\vec{\Omega}_C\cdot\vec{{\cal L}}]\vec{\Psi}^{(0)} ,
\label{eta-1}\\
{\partial\over{\partial t_{-1}}}\vec{\Psi}^{(1)}+
{\partial\over{\partial t_0}}\vec{\Psi}^{(0)}&=&[{\cal D}^*-
{1\over{4\tau_\perp}}{\cal L}^2
+i\vec{\Omega}_C\cdot\vec{{\cal L}}]\vec{\Psi}^{(1)}
+\left(\vec{\Omega}_{IA}+\vec{\Omega}_g\right)\times\vec{\Psi}^{(0)}~,
\label{eta0}\\
{\partial\over{\partial t_{-1}}}\vec{\Psi}^{(2)}+
{\partial\over{\partial t_0}}\vec{\Psi}^{(1)}+
{\partial\over{\partial t_1}}\vec{\Psi}^{(0)}&=&[{\cal D}^*-
{1\over{4\tau_\perp}}{\cal L}^2
+i\vec{\Omega}_C\cdot\vec{{\cal L}}]\vec{\Psi}^{(2)}
+\left(\vec{\Omega}_{IA}+\vec{\Omega}_g\right)\times\vec{\Psi}^{(1)}
\nonumber\\
&+&\vec{\Omega}_L\times\vec{\Psi}^{(0)}
-{\bf R}\vec{\Psi}^{(0)}~.
\label{eta+1}
\end{eqnarray}
For the analysis of this set of equations
it is convenient to split $\vec{\Psi}^{(n)}$ into an angle averaged and a
remaining part,
\begin{eqnarray}
\vec{\Psi}^{(n)}(\varepsilon,\omega,t_{-1},t_0,t_1)=
\vec{a}^{(n)}(\varepsilon,t_{-1},t_0,t_1)+
\delta\vec{a}^{(n)}(\varepsilon,\omega,t_{-1},t_0,t_1)~,
\label{splitting}
\end{eqnarray}
where $\vec{a}^{(n)}=<\vec{\Psi}^{(n)}>_\omega$, with an angle average defined by
\begin{eqnarray}
\langle(...)\rangle_\omega=\int d\omega (...)
\label{angleave}
\end{eqnarray}
and $<\delta\vec{a}^{(n)}>_\omega=0$ by definition.
Since the angle variables are periodic this partitioning
is always possible. From the initial condition,
$\vec{\Psi}(\varepsilon,\omega,0)=\vec{e}$, we
infer the intitial conditions
$\vec{a}^{(n)}(\varepsilon,0,0,0)=\vec{e}\delta_{n,0}$, and
$\delta\vec{a}^{(n)}(\varepsilon,\omega,0,0,0)=0$.
Recalling that the factor $J(\varepsilon)/(2\pi)^d n_s$ is included in the
definition of
$\delta\vec{S}(\varepsilon,\omega,t)$, the macroscopic non-equilibrium spin
polarization defined in Eq. (\ref{totaltest}) can now be rewritten as
\begin{eqnarray}
\delta\vec{S}(t)=\int_0^\infty d\varepsilon p(\varepsilon) \vec{a}(\varepsilon,t)
=\langle \vec{a}(\varepsilon,t) \rangle_\varepsilon~,
\label{Save1}
\end{eqnarray}
where we defined an energy average
\begin{eqnarray}
\langle(...)\rangle_\varepsilon=
\int_0^\infty d\varepsilon p(\varepsilon)(...)~.
\label{energyave}
\end{eqnarray}
Note that the function
$p(\varepsilon)$, which determines the terminating state of the fast,
spin-conserving
time evolution, enters here naturally as a weight function. Formally, the weight
function appears in our theory because of the Ansatz (\ref{deltaS}),
which enabled us to switch to the adjoint Fokker-Planck
equation. The expansion of $\vec{\Psi}$ implies an analogous expansion
for the macroscopic
($\vec{k}$-averaged) spin polarization:
\begin{eqnarray}
\delta\vec{S}(t)=\delta\vec{S}^{(0)}(t_{-1},t_0,t_1) +
\eta \delta \vec{S}^{(1)}(t_{-1},t_0,t_1)
+\eta^2 \delta \vec{S}^{(2)}(t_{-1},t_0,t_1)~.
\label{Save2}
\end{eqnarray}
We now calculate the leading order contribution $\delta\vec{S}^{(0)}(t)$
uniformly valid for all times. As a result, we will obtain a Bloch equation
which determines the long time behavior of the macroscopic spin
polarization.

\subsubsection{$O(\eta^{-1})$ equation}

With the substitution (\ref{splitting}), the $O(\eta^{-1})$ equation (\ref{eta-1}) splits
into two independent equations, one for the angle averaged part and one for the
angle dependent part.
\begin{eqnarray}
{\partial\over{\partial t_{-1}}}\vec{a}^{(0)}&=&{\cal D}^*\vec{a}^{(0)}~,
\label{zeroth1}\\
{\partial\over{\partial t_{-1}}}\delta \vec{a}^{(0)}&=&
[{\cal D}^*-{1\over{4\tau_\perp}}{\cal L}^2]\delta \vec{a}^{(0)}~.
\label{zeroth2}
\end{eqnarray}
Solutions of Eqs. (\ref{zeroth1}) and (\ref{zeroth2}) compatible with the
two initial conditions, $\vec{a}^{(0)}(\varepsilon,0,0,0)=\vec{e}$ and
$\delta \vec{a}^{(0)}(\varepsilon,0,0,0)=0$, are
$\vec{a}^{(0)}(\varepsilon,t_{-1},t_0,t_1)=\vec{a}^{(0)}(t_0,t_1)$,
with $\vec{a}^{(0)}(0,0)=\vec{e}$, and
$\delta \vec{a}^{(0)}(\varepsilon,\omega,t_{-1},t_0,t_1)=0$. Using
Eqs. (\ref{Save1}) and (\ref{Save2}), we find therefore
\begin{eqnarray}
\delta\vec{S}^{(0)}(t_0,t_1)=\vec{a}^{(0)}(t_0,t_1)~,
\label{S0}
\end{eqnarray}
that is, due to our choice of the initial condition, the macroscopic
zeroth order spin polarization is independent of the fast spin
conserving time scale $t_{-1}$ and solely evolves on the long time scales
$t_0$ and $t_1$.

\subsubsection{$O(\eta^{0})$ equation}

To determine the time evolution of the macroscopic spin polarization
on the long time scales $t_0$ and $t_1$, we study the $O(\eta^{0})$
equation. Substituting
(\ref{splitting}) into Eq. (\ref{eta0}), the latter splits into two independent
equations:
\begin{eqnarray}
{\partial\over{\partial t_{-1}}}\vec{a}^{(1)}&=&
-{\partial\over{\partial t_0}}\vec{a}^{(0)}+{\cal D}^*\vec{a}^{(1)}~,
\label{eta0ave}\\
{\partial\over{\partial t_{-1}}}\delta\vec{a}^{(1)}&=&
[{\cal D}^*-{1\over{4\tau_\perp}}
{\cal L}^2 +i\vec{\Omega}_C\cdot\vec{{\cal L}}]\delta\vec{a}^{(1)}
+\left(\vec{\Omega}_{IA}+\vec{\Omega}_g\right)\times\vec{a}^{(0)}~,
\label{eta0angle}
\end{eqnarray}
where we have used $\delta\vec{a}^{(0)}=0$, $\langle\vec{\Omega}_{IA}\rangle_\omega=0$,
and $\langle\vec{\Omega}_g\rangle_\omega=0$.
Applying the energy average $\langle (...) \rangle_\varepsilon$ on both sides of
Eq. (\ref{eta0ave}) yields
\begin{eqnarray}
{\partial\over{\partial t_{-1}}}\delta \vec{S}^{(1)}=
-{\partial\over{\partial t_0}}\delta \vec{S}^{(0)}
-\langle {\cal D}^*\vec{a}^{(1)}\rangle_\varepsilon~,
\label{eta0aveaux}
\end{eqnarray}
which, using in the second term partial integration and the definition of
$p(\varepsilon)$, reduces to
\begin{eqnarray}
{\partial\over{\partial t_{-1}}}\delta \vec{S}^{(1)}=
-{\partial\over{\partial t_0}}\delta \vec{S}^{(0)}~.
\label{eta0avefinal}
\end{eqnarray}
The vanishing of the second term on the rhs of Eq. (\ref{eta0aveaux}) is the result of
the spin conservation of the differential operator ${\cal D}$, which in turn is
ensured by the identity 
\begin{eqnarray}
{d\over{d\varepsilon}}{\varepsilon^2\over{\tau_d(\varepsilon)}}\bigg|_{\varepsilon=0}
={\varepsilon\over{\tau_f(\varepsilon)}}\bigg|_{\varepsilon=0}~.
\label{spinconservation}
\end{eqnarray}

It is crucial to note that the rhs of Eq. (\ref{eta0avefinal}) is independent of
the fast time $t_{-1}$, because $\vec{a}^{(0)}$ is independent of $t_{-1}$.
Integrating Eq. (\ref{eta0avefinal}) with respect to  $t_{-1}$, the rhs
therefore gives rise to a secular term, i.e. a term which is
proportional to $t_{-1}$. As a result, $\delta\vec{S}^{(1)}$
can be larger than $\delta\vec{S}^{(0)}$
for sufficiently large times. The expansion (\ref{Save2}) would be
valid only for short times, i.e. the expansion is non-uniform.
Within multiple time scale perturbation theory, secular terms can be
avoided by an appropriate choice of the time evolutions on the various time
scales. The secular term in Eq. (\ref{eta0avefinal}) can be particularly simply
removed by forcing the rhs to be zero, which gives rise to the condition
\begin{eqnarray}
\int_0^\infty d\varepsilon p(\varepsilon){\partial\over{\partial t_0}}
\vec{a}^{(0)}(t_0,t_1)=0~.
\label{EYlevel3}
\end{eqnarray}
That is, $\vec{a}^{(0)}(t_0,t_1)=\vec{a}^{(0)}(t_1)$, which, using
Eq. (\ref{S0}), leads to
$\delta\vec{S}^{(0)}(t_1)=\vec{a}^{(0)}(t_1)$, i.e. the time evolution
of the  zeroth order macroscopic spin polarization (and therefore the
spin decay) occurs solely on the long time scale $t_1$.
Since the rhs of Eq. (\ref{eta0avefinal}) is made to vanish, we
also obtain $\delta\vec{S}^{(1)}(t_{-1},t_0,t_1)=\delta\vec{S}^{(1)}(t_0,t_1)$, i.e.,
$\delta\vec{S}^{(1)}$ is independent of the fast time variable $t_{-1}$.
Using the definition of $\delta\vec{S}^{(1)}$, we furthermore conclude
that $\vec{a}^{(1)}(\varepsilon,t_{-1},t_0,t_1)=
\vec{a}^{(1)}(\varepsilon,t_0,t_1)$. Both results we need
in the analysis of the $O(\eta^{1})$ equations, which is necessary to
determine the time evolution on the remaining time scale $t_1$.

\subsubsection{$O(\eta^{1})$ equations}

To investigate the $O(\eta^1)$ equation (\ref{eta+1}), we substitute
(\ref{splitting}) into (\ref{eta+1}). Averaging over the angle and energy,
we find
\begin{eqnarray}
{\partial\over{\partial t_{-1}}}\delta\vec{S}^{(2)}=
-{\partial\over{\partial t_0}}\delta\vec{S}^{(1)}
-{\partial\over{\partial t_1}}\delta\vec{S}^{(0)}
+\langle \left(\vec{\Omega}_{IA}+\vec{\Omega}_g\right)
\times\delta\vec{a}^{(1)}\rangle_{\varepsilon,\omega}
+\vec{\Omega}_L\times\delta\vec{S}^{(0)}
-\langle {\bf R} \rangle_{\varepsilon,\omega}\delta\vec{S}^{(0)}~,
\label{eta+1aux}
\end{eqnarray}
where we used $\delta\vec{a}^{(0)}=0$,
$\langle\vec{\Omega}_{IA}\rangle_{\varepsilon,\omega}=0$,
$\langle\vec{\Omega}_g\rangle_{\varepsilon,\omega}=0$, and
$\langle{\cal D}^*\vec{a}^{(2)}\rangle_{\varepsilon,\omega}=0$. The third term
on the rhs of (\ref{eta+1aux}) contains $\delta\vec{a}^{(1)}$ which has to
be obtained from Eq. (\ref{eta0angle}), the angle dependent part of the
$O(\eta^0)$ equation. Before we proceed with the analysis of Eq. (\ref{eta+1aux})
let us therefore turn to Eq. (\ref{eta0angle}).
Integration of Eq. (\ref{eta0angle}) with respect
to the fast time $t_{-1}$ produces a secular term,
$(\vec{\Omega}_{IA}+\vec{\Omega}_g)\times\vec{a}^{(0)} t_{-1}$,
which cannot be removed, because
both $\vec{\Omega}_{IA}+\vec{\Omega}_g$ and
$\vec{a}^{(0)}$ are finite. A way to avoid
the resulting non-uniformity is to demand
\begin{eqnarray}
{\partial\over{\partial t_{-1}}}\delta\vec{a}^{(1)}=0~,
\label{quasistat}
\end{eqnarray}
i.e., to enforce
$\delta\vec{a}^{(1)}(\varepsilon,\omega,t_{-1},t_0,t_1)=
\delta\vec{a}^{(1)}(\varepsilon,\omega,t_0,t_1)$, which reduces
Eq. (\ref{eta0angle}) to
\begin{eqnarray}
[{\cal D}^*-{1\over{4\tau_\perp}}{\cal L}^2
+i\vec{\Omega}_C\cdot\vec{{\cal L}}]\delta\vec{a}^{(1)}
+\left(\vec{\Omega}_{IA}+\vec{\Omega}_g\right)\times\delta\vec{S}^{(0)}=0~,
\label{torque}
\end{eqnarray}
where we used in the last term $\vec{a}^{(0)}(t_1)=\delta\vec{S}^{(0)}(t_1)$.
The condition (\ref{quasistat}) is reminiscent of the quasi-stationarity assumption
usually invoked in the calculation of the D'yakonov-Perel' relaxation
rates.~\cite{Dyakonov72}
The multiple time scale approach enables us to identify the time scale on which
this assumption holds.

We now return to Eq. (\ref{eta+1aux}). Since $\delta\vec{a}^{(1)}$,
$\delta\vec{S}^{(1)}$, and
$\delta\vec{S}^{(0)}$ are independent of the fast time variable $t_{-1}$, the
whole rhs of Eq. (\ref{eta+1aux}) is independent of $t_{-1}$. Integration
with respect to $t_{-1}$ thus gives rise to a secular term which
has to be removed. We force therefore the rhs of Eq. (\ref{eta+1aux})
to vanish which can be certainly accomplished if we separately
demand
\begin{eqnarray}
{\partial\over{\partial t_0}}\delta\vec{S}^{(1)}&=&0~,
\\
{\partial\over{\partial t_1}}\delta\vec{S}^{(0)}&=&
\vec{\Omega}_L\times\delta\vec{S}^{(0)}
+\langle
\left( \vec{\Omega}_{IA}+\vec{\Omega}_g \right)
\times\delta\vec{a}^{(1)}\rangle_{\varepsilon,\omega}
-\langle {\bf R} \rangle_{\varepsilon,\omega}\delta\vec{S}^{(0)}~.
\label{Blochprecursor}
\end{eqnarray}
>From the first equation we find
\begin{eqnarray}
\int_0^\infty d\varepsilon p(\varepsilon) {\partial\over{\partial t_0}}
\vec{a}^{(1)}(\varepsilon,t_0,t_1)=0~,
\end{eqnarray}
that is $\vec{a}^{(1)}$ is independent of $t_0$, a result which we
need below. The second equation is already a precursor of the Bloch equation
for $\delta\vec{S}^{(0)}(t_1)$. Although it determines
$\delta\vec{S}^{(0)}(t_1)$ for a given $\delta\vec{a}^{(1)}(t_0,t_1)$, it
is however not yet a Bloch equation because, at this point of the calculation,
$\delta\vec{a}^{(1)}$ is still a function of $t_1$ and $t_0$.

To obtain a closed Bloch equation on the time scale $t_1$ alone,
we now examine the $t_0$ dependence of $\delta\vec{a}^{(1)}(t_0,t_1)$.
Towards that end, we consider
the angle dependent part of the $O(\eta^1)$ equation (\ref{eta+1}), which
reads
\begin{eqnarray}
{\partial\over{\partial t_{-1}}}\delta\vec{a}^{(2)}&=&
[{\cal D}^*-{1\over{4\tau_\perp}}{\cal L}^2
+i\vec{\Omega}_C\cdot\vec{{\cal L}}]\delta\vec{a}^{(2)}
\nonumber\\
&+&\left(\vec{\Omega}_{IA}+\vec{\Omega}_g\right)\times\vec{a}^{(1)}
+\left(\vec{\Omega}_{IA}+\vec{\Omega}_g\right)\times\delta\vec{a}^{(1)}
-\langle \left(\vec{\Omega}_{IA}+\vec{\Omega}_g\right)
\times\delta\vec{a}^{(1)} \rangle_{\omega}
\nonumber\\
&-&{\bf R}\vec{a}^{(0)}+\langle {\bf R}\rangle_{\omega} \vec{a}^{(0)}
-{\partial\over{\partial t_0}}\delta\vec{a}^{(1)}
\label{eta+1angle}~.
\end{eqnarray}
Except for the first term, all terms on the rhs are independent of $t_{-1}$,
and therefore give rise to secular terms. To remove the secular terms,
we set the undesired terms on the rhs to zero
\begin{eqnarray}
{\partial\over{\partial t_0}}\delta\vec{a}^{(1)}&=&
\left(\vec{\Omega}_{IA}+\vec{\Omega}_g\right)\times\delta\vec{a}^{(1)}
-\langle \left(\vec{\Omega}_{IA}+\vec{\Omega}_g\right)
\times\delta\vec{a}^{(1)} \rangle_{\omega}
\nonumber\\
&-&\large[{\bf R}-\langle {\bf R}\rangle_{\omega}\large]\vec{a}^{(0)}
+\left(\vec{\Omega}_{IA}+\vec{\Omega}_g\right)\times\vec{a}^{(1)}~.
\label{eta+1aux1}
\end{eqnarray}
The last two terms on the rhs of this equation are independent of
$t_0$ and therefore again give rise to a secular term if Eq.
(\ref{eta+1aux1}) is integrated with respect to $t_0$. In general,
the last two terms are finite. Thus, to avoid non-uniformity we
demand that $\delta\vec{a}^{(1)}$ is independent of $t_0$, that
is, we enforce $\delta\vec{a}^{(1)}(\varepsilon,\omega,t_0,t_1)=
\delta\vec{a}^{(1)}(\varepsilon,\omega,t_1)$.  With this
constraint, Eq. (\ref{eta+1aux1}) could be used to determine
$\vec{a}^{(1)}(\varepsilon,t_1)$ and eventually $\delta
S^{(1)}(t_1)$.

Because $\delta\vec{a}^{(1)}$ is independent of $t_0$, Eq. (\ref{Blochprecursor})
is in fact a Bloch equation on the time scale $t_1$ alone. The function
$\delta\vec{a}^{(1)}(\varepsilon,\omega,t_1)$ satisfies
Eq. (\ref{torque}), which, through $\delta\vec{S}^{(0)}(t_1)$, contains $t_1$
only as a parameter. Therefore, the function
$\delta\vec{a}^{(1)}(\varepsilon,\omega,t_1)$ instantaneously adjusts to
the function $\delta\vec{S}^{(0)}(t_1)$, which, in this sense, acts like a
``slave field'' for $\delta\vec{a}^{(1)}(\varepsilon,\omega,t_1)$.

To make the equations determining the decay of the macroscopic
spin polarization explicit, we recall $t_n=\eta^n t$ and go
back to the original, unscaled time variable and functions. As a result
Eq. (\ref{Blochprecursor}) becomes a Bloch equation for the macroscopic
spin polarization,
\begin{eqnarray}
{\partial\over{\partial t}}\delta\vec{S}^{(0)}(t)=
\vec{\Omega}_L\times\delta\vec{S}^{(0)}(t)
-\large[{\bf \Gamma}_{\rm EY}+{\bf \Gamma}_{\rm MN}\large]\delta\vec{S}^{(0)}(t)~,
\label{Bloch}
\end{eqnarray}
with initial condition $\delta\vec{S}(0)=\vec{e}$. The Elliott-Yafet
and motional-narrowing spin relaxation tensors are given by
\begin{eqnarray}
{\bf \Gamma}_{\rm EY}&=&\langle {\bf R} \rangle_{\varepsilon,\omega} ,
\\
{\bf \Gamma}_{\rm MN}\delta\vec{S}^{(0)}(t)&=&
-\langle \left(\vec{\Omega}_{IA}+\vec{\Omega}_g\right)
\times\delta\vec{a}^{(1)}\rangle_{\varepsilon,\omega}~,
\label{MNtensor}
\end{eqnarray}
respectively, and $\delta\vec{a}^{(1)}$ is obtained from Eq. (\ref{torque}),
which for convenience we state here again
\begin{eqnarray}
[{\cal D}^*-{1\over{4\tau_\perp}}{\cal L}^2 +
i\vec{\Omega}_C\cdot\vec{{\cal L}}]\delta\vec{a}^{(1)}
+\left(\vec{\Omega}_{IA}+\vec{\Omega}_g\right)\times\delta\vec{S}^{(0)}=0~.
\nonumber
\end{eqnarray}
Equations (\ref{Bloch}) -- (\ref{MNtensor}) are the main result of this
section.  They control the time evolution of the macroscopic spin polarization
on the long time scale, where spin relaxation, i.e., decay, occurs.
Equation (\ref{MNtensor}) is an implicit definition of the spin relaxation
tensor ${\bf \Gamma}_{\rm MN}$. The explicit form of ${\bf \Gamma}_{\rm MN}$
can be obtained by inserting the solution of Eq. (\ref{torque}), which is
always linear in $\delta\vec{S}(t)$, and
performing the angle and energy averages. In the next section we
illustrate this procedure for a doped quantum well subject to a small magnetic
field.

The macroscopic spin relaxation tensor contains the EY process and the
motional-narrowing (DP and VG) processes.
Due to the different angle dependences of the two main motional-narrowing
spin relaxation processes, DP and VG processes, Eq. (\ref{MNtensor})
splits for isotropic semiconductors into two separate terms,
${\bf \Gamma}_{\rm DP}$ and ${\bf \Gamma}_{\rm VG}$. Accordingly, for
isotropic semiconductors, a Matthiessen-type rule holds for the total
spin relaxation tensor,
${\bf \Gamma}={\bf \Gamma}_{\rm EY}+
{\bf \Gamma}_{\rm DP}+{\bf \Gamma}_{\rm VG}$, and,
as a consequence, for the spin relaxation rates,
which are the diagonal elements of the relaxation tensors.~\cite{our}
The quenching of the motional-narrowing processes due
to the orbital motion is contained in Eq. (\ref{torque})
through the term proportional to $\vec{\Omega}_C$.~\cite{Ivchenko73,our}
The on- and off-shell relaxation rates
appearing in Eq. (\ref{torque}) are total relaxation rates due to
whatever scattering processes are included. A Matthiessen rule holds
separately for the on- and off-shell rates. The energy average
$\langle(...)\rangle_{\varepsilon}$ is defined in Eq. (\ref{energyave}).
Most importantly, it contains a weight function $p(\varepsilon)$
defined as the solution of Eq. (\ref{pdiff}).
This function describes the energy dependence of the quasi-stationary
spin polarization which appears on the short time scale because of
fast, spin-conserving inelastic scattering processes, i.e. because
of energy relaxation and diffusion. Once the quasi-stationary
spin polarization is established, it slowly decays on the long time
scale set by the spin non-conserving terms in the Fokker-Planck equation.

\section{Application to quantum well structures}

In this section we apply the formalism to
a quantum well at low enough temperatures, where electron-impurity and
electron-electron scattering dominate. We are here particularly interested
in the effects of Pauli blocking and inelasticity. For illustration, we
focus therefore only
on the DP process, which, for small to moderate magnetic fields, is usually
the dominant spin relaxation process. Moreover, if the magnetic
field is small enough, the time $\hat{\tau}_C$ it takes to complete a cyclotron 
orbit is much longer than any of the intrinsic scattering times $\hat{\tau}_i$
($i = f, d, \perp$), and the quenching effect of the magnetic
field can be ignored.

We consider a symmetric GaAs quantum well, grown in the $[001]$ direction,
which is also the quantization axis for the electron spin.
Due to the assumed structural symmetry, there is only bulk inversion asymmetry
giving
rise to DP spin relaxation.~\cite{Dyakonov86} As in
the bulk case, we treat the two states at the conduction band
minimum explicitly and include a large set of states
perturbatively, up to third order, to include the effect of bulk
inversion asymmetry.
For energies close to the band minimum, the
Hamiltonian for the quantum well can be cast into the form (\ref{model}).
The spin off-diagonal term in the quantum well Hamiltonian is
the bulk spin off-diagonal term averaged over the envelope function
of the conduction subband. Assuming for simplicity infinite
confinement and restricting the calculation to the lowest
conduction subband, we find (neglecting cubic terms in $\vec{k}$)
\begin{eqnarray}
\hbar\vec{\Omega}^{QW}_{IA}(\varepsilon,\phi)&=&2\delta_0 \sqrt{\varepsilon}({\pi\over
L})^2\left(\begin{array}{c}
-\cos{\phi}\\
\sin{\phi}\\
0
\end{array}\right)
\nonumber\\
&=&{1\over{\tau_{IA}(\varepsilon)}}\vec{\kappa}_{IA}(\phi)~,
\label{precession}
\end{eqnarray}
where we have defined a precession rate 
$1/\tau_{IA}(\varepsilon)=C_{IA}^{QW}\sqrt{\varepsilon}$ with
$C_{IA}^{QW}=2\delta_0(\pi/L)^2$.

Since we are only interested in the DP spin relaxation tensor, we
neglect in Eq. (\ref{torque})
the torque force due to $\vec{\Omega}_g$. Because we furthermore assume small
magnetic fields, we also ignore the orbital motion of the electrons.
The separation Ansatz,
\begin{eqnarray}
\delta\vec{a}^{(1)}(\varepsilon,\phi,t)=
\tau(\varepsilon)\vec{\kappa}_{IA}(\phi)\times\vec{S}^{(0)}(t)~,
\label{deltaa1}
\end{eqnarray}
then reduces Eq. (\ref{torque}) to a scalar differential equation,
\begin{eqnarray}
[{\cal D}^*-{1\over{4\tau_\perp(\varepsilon)}}]\tau(\varepsilon)
+{1\over{\tau_{IA}(\varepsilon)}}=0~,
\label{taudiffeq}
\end{eqnarray}
which determines the generalized relaxation time
$\tau(\varepsilon)$. Because the differential operator ${\cal
D}^*$ accounts for inelastic scattering, we have to conclude that
even on the long time scale, where the spin polarization decays,
inelasticity cannot be ignored. Thus, inelastic scattering
processes not only determine the initial condition for the decay
stage but they directly affect the time evolution (of the
macroscopic spin polarization) in the decay stage.  Multiplying
from the left Eq. (\ref{taudiffeq}) by $p(\varepsilon)\tau(\varepsilon)$,
integrating the resulting equation over $\varepsilon$, and using
condition (\ref{spinconservation}) yields an equivalent differential 
equation,
\begin{eqnarray}
[{\cal D}-{1\over{4\tau_\perp(\varepsilon)}}]p(\varepsilon)\tau(\varepsilon)
+{{p(\varepsilon)}\over{\tau_{IA}(\varepsilon)}}=0~,
\label{taudiffeq1}
\end{eqnarray}
which can be also used to determine $\tau(\varepsilon)$.

Inserting Eq. (\ref{deltaa1}) into (\ref{MNtensor}), ignoring the 
$\Omega_g$ term, and performing the angle and energy 
averages finally yields for the DP spin relaxation tensor
\begin{eqnarray}
{\bf \Gamma}_{\rm DP}={1\over\tau_{\rm DP}}
\left( \begin{array}{ccc}
1 & 0 & 0 \\
0 & 1 & 0 \\
0 & 0 & 2
\end{array}
\label{tensor}
\right)~,
\end{eqnarray}
with the DP spin relaxation rate given by
\begin{eqnarray}
{1\over\tau_{\rm DP}}=\pi\langle {\tau\over{\tau_{IA}}}\rangle_\varepsilon~,
\label{DPrate}
\end{eqnarray}
where the energy average is defined in Eq. (\ref{energyave}).
To determine the function $p(\varepsilon)$, we integrate Eq. (\ref{pdiff}),
which gives
\begin{eqnarray}
\left[-v(\varepsilon)+\frac{d}{d\varepsilon}w(\varepsilon)\right]
p(\varepsilon)=0~,
\end{eqnarray}
where we used again condition (\ref{spinconservation}) and 
introduced the dynamical friction and diffusion coefficients in 
$\varepsilon$-space, $v(\varepsilon)=\varepsilon/\tau_f(\varepsilon)$ and 
$w(\varepsilon)=\varepsilon^2/\tau_d(\varepsilon)$, respectively. 
Integrating once more, we obtain
\begin{eqnarray}
p(\varepsilon)=p(0)e^{\int_0^\varepsilon d\varepsilon'
\frac{v(\varepsilon')-w'(\varepsilon')}{w(\varepsilon')}}~,
\label{p}
\end{eqnarray}
with $w'(\varepsilon)=dw(\varepsilon)/d\varepsilon$ and a normalization
constant $p(0)$, which we fix according to
\begin{eqnarray}
\int_0^{2\pi}d\phi\int_0^\infty d\varepsilon p(\varepsilon)=1~.
\label{norm}
\end{eqnarray}
Note, because of condition (\ref{spinconservation}), the integral in Eq. (\ref{p})
is zero for $\varepsilon\rightarrow 0$. That is, $p(\varepsilon)$ is
well-defined for $\varepsilon\rightarrow 0$.

The Bloch equation (\ref{Bloch}) for the macroscopic spin polarization has to
be solved with the spin relaxation tensor (\ref{tensor}) and taking
the particular geometry of the experimental set-up
into account. Here, we consider the case of Kerr or Faraday rotation
experiments, where the small magnetic field, which causes the spin
precession, is along the x-axis. The propagation direction of the
pump and probe pulses is assumed to be perpendicular to the quantum well
plane, i.e., parallel to the z-axis (growth axis). The initial spin polarization
is therefore along the z-axis, i.e. $\vec{e}=(0,0,1)^T$, and the probe pulse
monitors the decay of a spin polarization which precesses in the yz-plane. Note
that the spin decay in the yz-plane plane is not isotropic
($\Gamma_{\rm yy}\neq \Gamma_{\rm zz}$). Assuming $\tau_{\rm DP}\gg 1/\Omega_L$,
the solution of the Bloch equation is
\begin{eqnarray}
\delta\vec{S}(t)=
\left(\begin{array}{c}
0\\
-\sin{\Omega_L t}
\\
\cos{\Omega_L t}
\end{array}\right)e^{-\Gamma t},
\end{eqnarray}
where the decay rate of the spin polarization is given by
the arithmetic mean of the decay rates in y- and in z-direction:
$\Gamma=(\Gamma_{\rm yy}+\Gamma_{\rm zz})/2=3/(2\tau_{\rm DP})$.

\begin{figure}[t]
\hspace{0.0cm}\psfig{figure=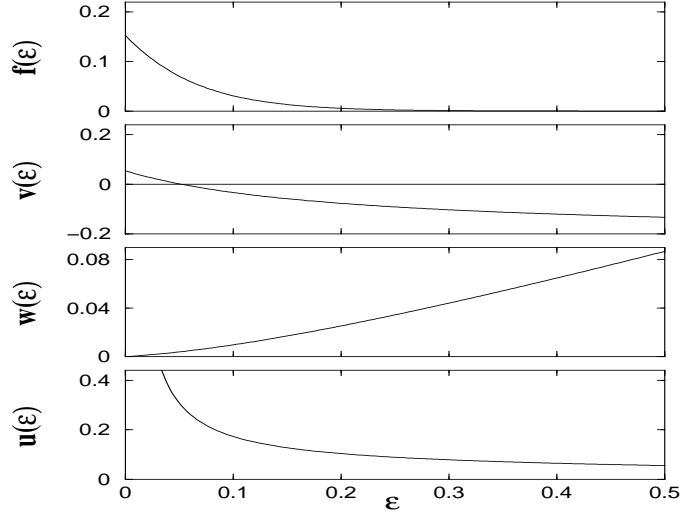,height=7.0cm,width=0.5\linewidth,angle=0}
\caption[fig1]
{Distribution function (for the equilibrated state at
$t\rightarrow\infty$), dynamical friction coefficient,
dynamical diffusion coefficient, and angle randomization
coefficient (from top to bottom) for a modulation doped quantum well at an
electron density $n=4\times10^9{\rm cm}^{-2}$ and $T=10{\rm K}$.
For the calculation of the coefficients, only electron-electron
scattering is taken into account. The unit of $\varepsilon$ is
${14.93~\rm meV}$. The units of the functions $v$, $w$, and $u$,
are ${22.68~\rm meV/ps}$, $338.44~{\rm (meV)^2/ps}$, and
$1.52~{\rm 1/ps}$, respectively. }
\label{fig1}
\end{figure}

The results presented below are for a $L=25{\rm nm}$ GaAs quantum
well. The parameter needed to specify
$\hbar\vec{\Omega}^{QW}_{IA}(\vec{k})$
is $\delta_0=0.06\hbar^3/\sqrt{(2m^*)^3\epsilon_g}$.~\cite{Ioffe} The
remaining parameters, such as the effective CB electron mass or the
static dielectric constant $\epsilon_b$ (needed for the Coulomb matrix
element) can be found in standard
data bases.\cite{parameter} Numerically, we first calculate
$v(\varepsilon)=\varepsilon/\tau_f(\varepsilon)$,
$w(\varepsilon)=\varepsilon^2/\tau_d(\varepsilon)$,
which define the differential operator ${\cal D}$,
and the on-shell rate $u(\varepsilon)=1/4\tau_\perp(\varepsilon)$ taking
electron-electron
and electron-impurity scattering into account (see Appendix). We then
determine $p(\varepsilon)$ from Eqs. (\ref{p}) and (\ref{norm}).
Finally, we solve the differential equation (\ref{taudiffeq1}) for
$\tau(\varepsilon)$ numerically and obtain the DP relaxation rate
$1/\tau_{\rm DP}$ by numerically integrating Eq. (\ref{DPrate}).

\begin{figure}[t]
\hspace{0.0cm}\psfig{figure=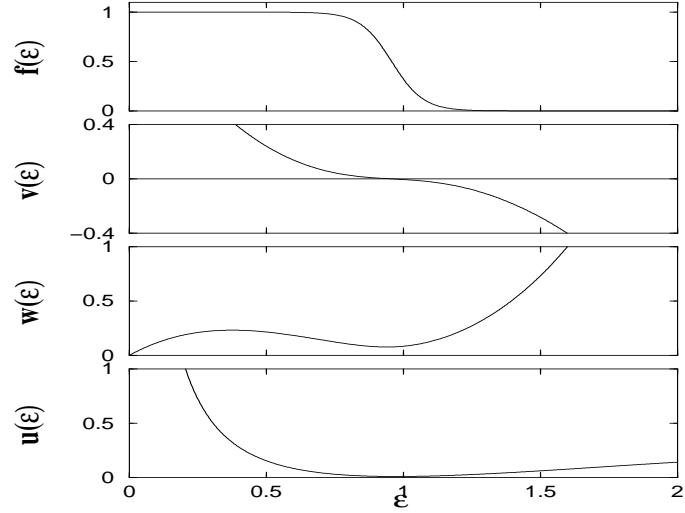,height=7.0cm,width=0.5\linewidth,angle=0}
\caption[fig2]
{Distribution function (for the equilibrated state at
$t\rightarrow\infty$), dynamical friction coefficient,
dynamical diffusion coefficient, and angle randomization
coefficient (from top to bottom) for a modulation doped quantum well at an
electron density $n=4\times10^{11}{\rm cm}^{-2}$ and $T=10{\rm K}$.
As in Fig. \ref{fig1}, for the calculation of the coefficients,
only electron-electron scattering is considered.  The units of
$\varepsilon$, $v$, $w$, and $u$ are the same as in Fig. \ref{fig1}.}
\label{fig2}
\end{figure}

In Figs. \ref{fig1} and \ref{fig2} we show, for $T=10{\rm K}$,
the dimensionless functions
$v(\varepsilon)=\varepsilon/\tau_f(\varepsilon)$,
$w(\varepsilon)=\varepsilon^2/\tau_d(\varepsilon)$, and
$u(\varepsilon)=1/4\tau_\perp(\varepsilon)$ for a modulation doped
quantum well with electron density $n=4\times10^9{\rm cm}^{-2}$
(non-degenerate electrons)
and $n=4\times10^{11}{\rm cm}^{-2}$ (degenerate electrons), 
respectively. In a modulation
doped quantum well electron-impurity scattering is negligible
because of the spatial separation between the dopants and the
electrons, we take therefore only electron-electron scattering
into account. The physical content of the functions
$v(\varepsilon)$ and $w(\varepsilon)$ is that of dynamical
friction and diffusion coefficients (in $\varepsilon$-space) for 
the ``test'' spin polarization resulting from the scattering between the
spin-polarized ``test'' electrons
and the equilibrated ``field'' electrons. The
function $u(\varepsilon)$ denotes the on-shell scattering rate
arising from the ``test'' electron's elastic scattering off
``field'' electrons. It randomizes the angle $\phi$;
hence $u(\varepsilon)$ can be interpreted as an angle
randomization coefficient. 

In Figs. \ref{fig1} and \ref{fig2}
we also show the electron distribution functions which characterize
the equilibrated state at $t\rightarrow\infty$.   
Because of our assumption that only a small portion of the total number
of electrons initially contributed to the ``test'' spin polarization
(i.e. $\delta f(\vec{k},t)\ll f(\vec{k})$), the equilibrium distribution 
functions are used to determine the friction, 
diffusion, and angle randomization coefficients (see section III.B). 
In other words, we approximated at $t<\infty$ the 
distributions of the friction and diffusion causing spin-balanced 
``field'' electrons by the distributions of the equilibrated 
electrons at $t\rightarrow\infty$, despite the fact that at $t<\infty$
of course a small number 
of electrons is still out of equilibrium and contributes to the finite spin 
polarization.  
Obviously, this is only permissible close enough to equilibrium.

Inelastic scattering encoded in $v(\varepsilon)=\varepsilon/\tau_f(\varepsilon)$
and $w(\varepsilon)=\varepsilon^2/\tau_d(\varepsilon)$
gives rise to spin-conserving energy relaxation and diffusion,
which in turn yields a quasi-stationary spin polarization
$\delta\vec{S}_{st}(\varepsilon,\phi)=p(\varepsilon)\vec{e}$, towards
which any initial spin polarization relaxes very quickly without losing spins.
The time scale on which this spin-conserving relaxation takes place is set by
the relaxation times $\tau_i(\varepsilon)$, $i = f, d, \perp$.
In Figure \ref{fig3} we show $p(\varepsilon)$
for $T=10{\rm K}$ and three densities: $n=4\times10^9{\rm cm}^{-2}$,
$n=4\times10^{11}{\rm cm}^{-2}$, and  $n=4\times10^{12}{\rm cm}^{-2}$. In the inset
we again depict the corresponding distribution functions for the electrons
which characterize the equilibrated state at $t\rightarrow\infty$.
At very low densities, where the electrons are non-degenerate, $p(\varepsilon)$
is centered around $\varepsilon=0$, while at high densities, where
the electrons are degenerate, $p(\varepsilon)$ is centered
around the Fermi energy for the electrons. Note, however, that 
$p(\varepsilon)$ describes the spin polarization at $t=0$ while 
$f(\varepsilon)$ is the equilibrium distribution of the electrons at 
$t\rightarrow\infty$. 

\begin{figure}[t]
\hspace{0.0cm}\psfig{figure=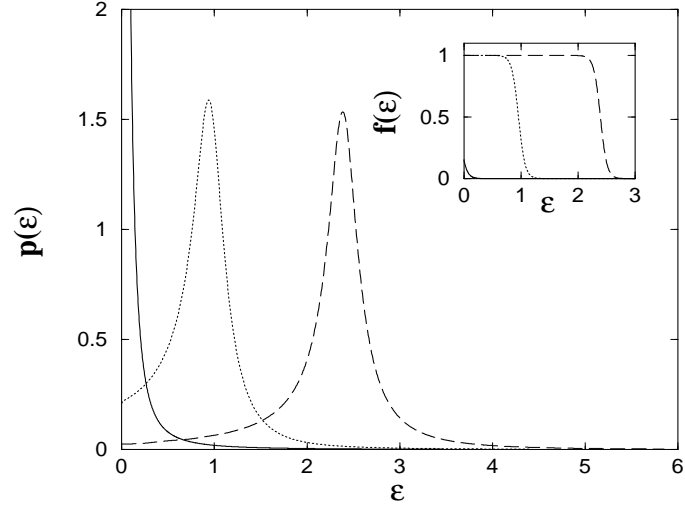,height=7.0cm,width=0.5\linewidth,angle=0}
\caption[fig3]
{Weight function for a modulation doped quantum well at $T=10{\rm K}$
and three electron densities $n=4\times10^9{\rm cm}^{-2}$ (solid
line), $n=4\times10^{11}{\rm cm}^{-2}$ (dotted line), $n=10^{12}{\rm cm}^{-2}$
(dashed line), taking only electron-electron scattering into account.
The inset shows the corresponding electron distribution functions
in the equilibrated state at $t\rightarrow\infty$.}
\label{fig3}
\end{figure}      

At this point, a brief discussion about the applicability of the diffusion 
approximation for the linearized collision integral $J_{ee}^{(0)}$ is in 
order.  The diffusion approximation is expected to be applicable because 
Eq. (\ref{CI0}) has the form of a master equation and can thus be formally 
expanded with 
respect to the momentum transfer $\vec{q}$. For a sufficiently rapidly 
decaying transition probability $W^{ee}(\vec{k};\vec{q})$, the expansion
can then be truncated after the second order term as in Eq. (\ref{diffapp}). 
The validity of the diffusion approximation depends therefore on the 
transition probability, which in dimensionless form reads 
[see Eq. (\ref{AppWee5}) in the Appendix] 
\begin{eqnarray}
W^{ee}(\vec{k};\vec{q})=\int^\infty_{-\infty}d\omega
|M(q)|^2[f(k^2-\omega)+n(-\omega)]{\rm Im}\tilde{\chi}(q,\omega)
\delta(\vec{k} \cdot \vec{q}-\frac{q^2}{2}-
\frac{\omega}{2})~,
\label{TransProb}
\end{eqnarray}
with $M(q)=1/(q+q_s)$ the Coulomb matrix element, $q_s$ the 
Thomas-Fermi screening wave number, and 
$f(x)$ and $n(x)$ the Fermi and Bose functions, respectively.  
Note, in Eq. (\ref{TransProb}) we do not expand the
distribution functions and energies. As a result, 
cut-off problems are avoided and the friction and diffusion coefficients, 
$A^{ee}_i(\vec{k})$ and $B^{ee}_{ij}(\vec{k})$ respectively,   
can be calculated without restricting the $\vec{q}$-integration. Thus, 
the diffusion approximation does not ignore hard scattering processes 
with large momentum transfer; it only treats them approximately,
whereas soft scattering processes are treated exactly. Moreover, from 
Eq. (\ref{TransProb}) it follows that $J_{ee}^{(0)}$ 
is actually a ``phonon-type'' collision integral, the role of phonons 
being played by the collective excitations of the spin-balanced 
``field'' electrons. Indeed, replacing $M(q)$ by the electron-phonon
matrix element and ${\rm Im}\tilde{\chi}(q,\omega)$ by the phonon 
spectral function, Eq. (\ref{TransProb}) gives the  
transition probability for spin-polarized ``test'' electrons 
scattering off equilibrated phonons. This analogy already suggests that 
a diffusion approximation is applicable to $J_{ee}^{(0)}$.
\begin{figure}[t]
\hspace{0.0cm}\psfig{figure=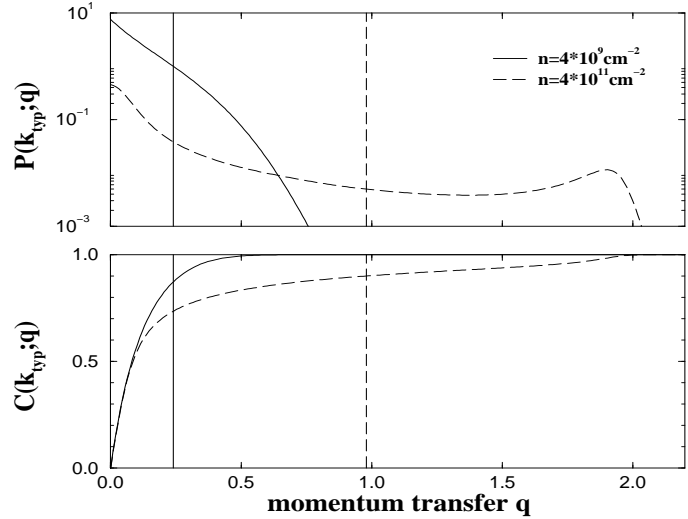,height=7.0cm,width=0.5\linewidth,angle=0}
\caption[FigDiff]
{
The upper panel shows $P(k_{\rm typ};q)$ for T=10K and
two electron densities: $n=4\times 10^9 {\rm cm}^{-2}$
(non-degenerate electrons) and $n=4\times 10^{11} {\rm cm}^{-2}$
(degenerate electrons). For non-degenerate electrons, the
typical momentum $k_{\rm typ}$ corresponds 
to the thermal energy (vertical solid line), whereas for degenerate 
electrons, $k_{\rm typ}$ is the Fermi wave number (vertical dashed line).
In the lower panel we present the cumulant $C(k_{\rm typ};q)$, 
defined in Eq. (\ref{cumulant}), which is a measure of the
relative importance of the scattering processes with a momentum transfer 
less than $q$.
}
\label{FigDiff}
\end{figure}           
To demonstrate its validity, it is however necessary to 
show that for typical 
values of $\vec{k}$ the transition probability $W^{ee}(\vec{k};\vec{q})$
indeed decays sufficiently rapidly for large $\vec{q}$. 

For that purpose we investigate   
\begin{eqnarray}
P(k_{\rm typ};q)=\int d\phi~q~W^{ee}(\vec{k}_{\rm typ};\vec{q})~,   
\end{eqnarray}
which is essentially the angle-averaged zeroth order moment of 
$W^{ee}(\vec{k}=\vec{k}_{\rm typ};\vec{q})$ with $\phi$ the angle between
$\vec{k}_{\rm typ}$ and $\vec{q}$. [The first and second moments 
appear in the calculation of the friction and diffusion coefficients 
$A^{ee}_i(\vec{k})$ and $B^{ee}_{ij}(\vec{k})$.] For degenerate electrons 
$k_{\rm typ}$ is the Fermi wave number $k_F$ while for non-degenerate 
electrons $k_{\rm typ}$ is the wave number corresponding to the 
thermal energy. In Fig. \ref{FigDiff} we show for the two parameter 
sets used in Figs. \ref{fig1} and \ref{fig2}, respectively, the zeroth 
order moment $P(k_{\rm typ};q)$ together with the cumulant 
\begin{eqnarray} 
C(k_{\rm typ};q)=\frac{\int_0^q dr P(k_{\rm typ};r)}
{\int_0^\infty dr P(k_{\rm typ};r)},  
\label{cumulant}
\end{eqnarray}
from which we can estimate the relative importance of scattering
processes with momentum transfer up to $q$. In the upper panel 
of  Fig. \ref{FigDiff} we see that $P(k_{\rm typ};q)$ indeed decays 
with increasing $q$. In the degenerate regime $q=2k_F$ scattering is 
clearly visible, but one order of magnitude less probable than the 
$q\rightarrow 0$ scattering process. From the cumulants, displayed 
in the lower panel, we infer moreover that in the non-degenerate as 
well as degenerate regime, soft scattering processes with $q \le 
k_{\rm typ}$ give the main contribution (around 80-90 $\%$).  In 
both cases we expect therefore the diffusion approximation to produce 
reasonable quantitative results for the DP spin relaxation time. To 
estimate the error precisely is complicated. It would require a 
detailed investigation of the full momentum dependence of 
$W^{ee}(\vec{k};\vec{q})$ together with a reference calculation 
which does not invoke the diffusion approximation. The spin 
relaxation times we obtain as a function of temperature and density 
compare favorably with experimental results, indicating that the 
modeling of scattering processes within a diffusion approximation 
is sufficient for the calculation of DP spin relaxation times.

\begin{figure}[t]
\hspace{0.0cm}\psfig{figure=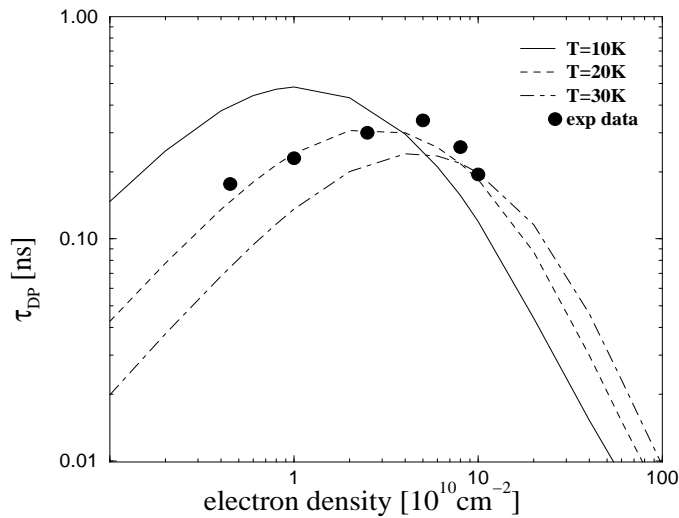,height=7.0cm,width=0.5\linewidth,angle=0}
\caption[fig4]
{D'yakonov-Perel' spin lifetime due to electron-electron scattering
for a $25 {\rm nm}$ modulation doped quantum well
as a function of electron density at three temperatures
$T=10{\rm K}, T=20{\rm K}, {\rm and}~T=30{\rm K}$. The solid dots
are experimental data at $T=10{\rm K}$ from Ref.~\cite{Sandhu01}. }
\label{fig4}
\end{figure}     

We now turn to the numerical results for the DP spin lifetime
$\tau_{DP}$. Figure \ref{fig4} shows $\tau_{DP}$ as a function of
electron density for a modulation doped quantum well at three
temperatures $T=10{\rm K}$, $20{\rm K},$ and $30{\rm K}$. Since in
a modulation doped quantum well electron-impurity scattering is
negligible, we take only electron-electron scattering into account.
For fixed temperature, the spin
lifetime first increases with electron density, reaches a maximum,
and then decreases again. The non-monotonic density dependence of the
spin lifetime follows the density dependence of the electron-electron
scattering rate. At low densities, the scattering rate is small because
of lack of scattering partners, while at high densities the scattering
rate is suppressed because of efficient Pauli blocking. At intermediate
densities, where the cross-over from non-degenerate to degenerate
electrons occurs, the electron-electron scattering rate, and thus the
DP spin lifetime, is maximal.
The position of the maximum shifts with decreasing temperature to
lower densities because the density, where the cross-over from a non-degenerate
to a degenerate electron gas takes place, decreases with temperature. The
relaxation time of photo-currents in optically pumped semiconductors shows a
similar non-monotonic density dependence.~\cite{Combescot87}

For a fixed electron density, the DP spin lifetime decreases with temperature
in the low density regime and increases with temperature in the high
density regime. The latter is because of the temperature induced reduction
of the Pauli blocking, giving rise to an increasing electron-electron
scattering rate and therefore to an increasing DP spin lifetime. In the
low density regime, on the other hand, increasing temperature broadens
the electron distribution function, i.e. electrons occupy states higher
up in the band. The average thermal energy increases therefore and the
spin decay occurs preferentially from states higher up in the band, where
the torque force induced by the bulk inversion asymmetry is larger.
As a consequence, the DP spin lifetime decreases with temperature in
the low density regime.

In Fig. \ref{fig4} we also plot experimental
data for $T=10{\rm K}$ from Ref.~\cite{Sandhu01}. For electron
densities above $n=5\times10^{10}{\rm cm}^{-2}$ the agreement between
theory and experiment is quite good, given the fact that our calculation
is based on an idealized quantum well with infinitely high confinement
potential. In this density regime, we expect our results to even
underestimate the spin lifetimes,
because the model for the electronic structure of the quantum well
most probably overestimates $\hbar\vec{\Omega}_{IA}^{QW}$. Indeed,
the constant defining the magnitude of the splitting of the conduction
subband $C^{QW}_{IA}\sim(\pi/L)^2\sim E_1^\infty$, with $E_1^\infty$ the
confinement energy of the lowest conduction subband. For a finite confinement
potential, $E_1$ is smaller than $E_1^{\infty}=\hbar^2\pi^2/2m^*L^2$,
giving rise to a smaller splitting and, consequently, to a larger $\tau_{DP}$.
To obtain in this density regime better agreement between experimentally measured
and theoretically calculated spin lifetimes an improved electronic
structure calculation is clearly necessary.
At lower densities, on the other hand, electrons
are most likely localized to donors (at $T=10{\rm K}$ thermal ionization is
negligibly small) and our theory, which is based on
a band picture, does not apply.

\begin{figure}[t]
\hspace{0.0cm}\psfig{figure=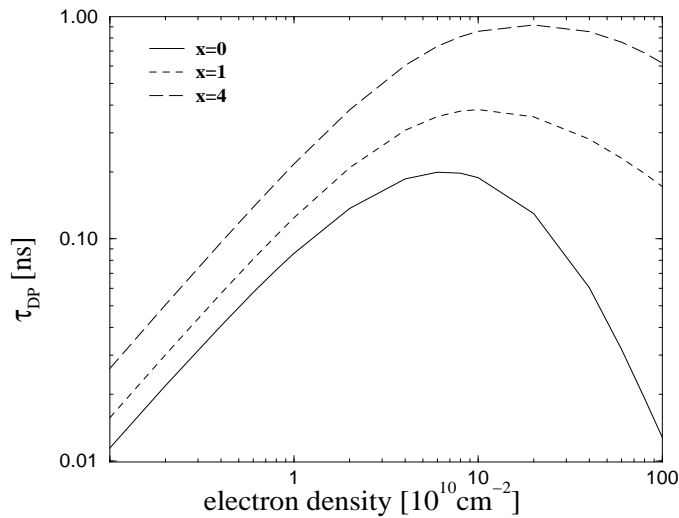,height=7.0cm,width=0.5\linewidth,angle=0}
\caption[fig5]
{D'yakonov-Perel' spin lifetime for a $25{\rm nm}$ quantum well
at $T=40{\rm K}$ and three values of ionized (donor and acceptor)
impurity concentrations: no impurities ($x=0$), ionized impurity
concentration equal to the electron density ($x=1$), and
ionized impurity concentration equal to four times the
electron density ($x=4$). For $x=0$, only electron-electron scattering
contributes to the spin lifetime, whereas for $x\neq 0$ both
electron-electron and electron-impurity scattering determine
the spin lifetime. }
\label{fig5}
\end{figure}

In quantum wells which are not modulation doped, electron-impurity
scattering due to donors and acceptors (in compensated samples) provides
an additional, very efficient scattering process. The DP spin lifetime
increases with scattering rate. As a result, we expect the spin
lifetimes in quantum wells that are not modulation doped to be substantially
longer than in modulation doped quantum wells. This can be seen in
Fig. \ref{fig5}, where we plot the electron density dependence of the
DP spin lifetime at $T=40{\rm K}$ for $x=0$ (modulation
doped), $x=1$ (uncompensated quantum well with equal impurity and
electron density), and $x=4$ (compensated quantum well with
impurity (donor and acceptor) density four times the electron
density). As expected, the spin lifetimes increase with $x$ for
all electron densities. The increase is however not uniform, with
the largest increase taking place at high electron densities,
where Pauli blocking very effectively suppressed the DP spin lifetime
in the modulation doped quantum well. The electron-impurity scattering
rate is not affected by Pauli blocking and leads therefore to a
substantial enhancement of the DP spin lifetime at high doping
levels.

\begin{figure}[t]
\hspace{0.0cm}\psfig{figure=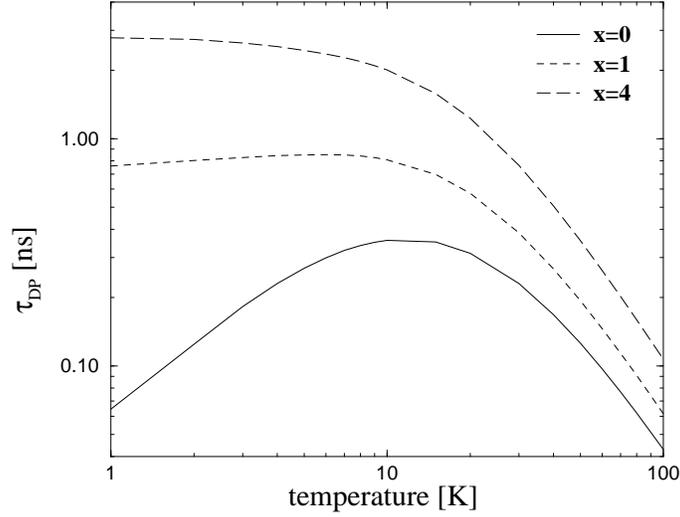,height=7.0cm,width=0.5\linewidth,angle=0}
\caption[fig6]
{D'yakonov-Perel' spin lifetime for a $25{\rm nm}$ quantum well as a function
of temperature at an electron density $n=3\times10^{10}{\rm cm}^{-2}$
and three values of ionized (donor and acceptor)
impurity concentrations: no impurities ($x=0$), ionized impurity
concentration equal to the electron density ($x=1$), and
ionized impurity concentration equal to four times the
electron density ($x=4$). As in Fig. \ref{fig5}, for $x=0$, only
electron-electron scattering contributes to the spin lifetime,
whereas for $x\neq 0$ both
electron-electron and electron-impurity scattering determine
the spin lifetime.}
\label{fig6}
\end{figure}

For a fixed density the character of the electron gas also changes with
temperature. In particular, increasing temperature pushes the electron
gas into the non-degenerate regime. For some temperature, the
cross-over from a degenerate to a non-degenerate electron gas occurs,
electron-electron scattering is particularly strong, and we expect
spin lifetimes to be enhanced.~\cite{Glazov03b} This effect is demonstrated
in Fig. \ref{fig6},
where we plot for an electron density $n=3\times10^{10}{\rm cm}^{-2}$ the
temperature dependence of the spin lifetime for $x=0, 1,$ and $4$.
In the modulation doped case ($x=0$) the enhancement of the
spin lifetime in the temperature range where the cross-over from
degenerate to non-degenerate takes place can be most clearly seen.
For finite $x$, the spin lifetimes are for all temperatures longer
than for $x=0$. The enhancement is again not uniform. At high
temperatures it is very small. While at low temperatures it is
very large, because in that range the Pauli blocking leads to a
strong suppression of the DP spin relaxation rate in modulation
doped quantum wells. In fact, the spin lifetimes at low temperatures
saturate by a value set by the electron-impurity scattering rate.
The maximum in the spin lifetime is therefore less pronounced 
(or even disappears completely) in
quantum wells that are not modulation doped.


\begin{figure}[t]
\hspace{0.0cm}\psfig{figure=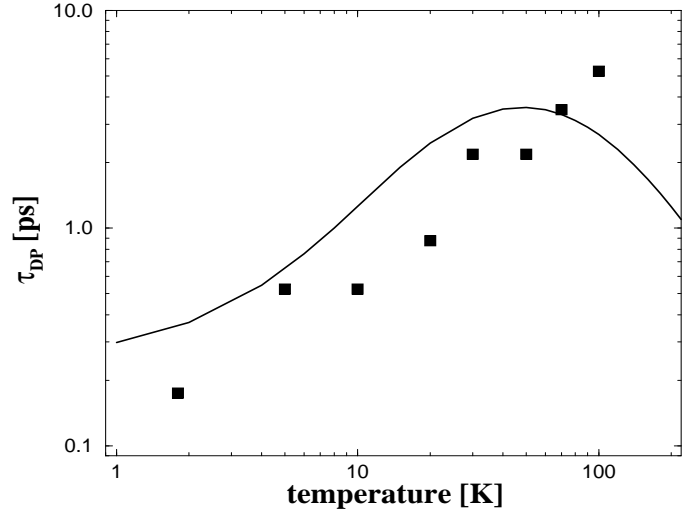,height=7.0cm,width=0.5\linewidth,angle=0}
\caption[fig7]
{D'yakonov-Perel' spin lifetime for a 10nm modulation doped GaAs 
quantum well as a function of temperature at an electron density 
$n=1.86\times10^{11}{\rm cm}^{-2}$ (solid line).  
The solid boxes are experimental data from Ref.~\cite{Brand02}.
}
\label{fig7}
\end{figure}          

In Fig. \ref{fig7}, we finally compare for a 10nm modulation doped 
GaAs quantum well with
an electron density $n=1.86\times10^{11}{\rm cm}^{-2}$ the theoretically
obtained temperature dependence of the D'yakonov-Perel' spin lifetime
with the experimentally measured temperature
dependence of the D'yakonov-Perel' spin lifetime.~\cite{Brand02}
Below 50K the agreement between the experimental data points and the
theoretical results is quite reasonable, suggesting that
in this temperature range electron-electron scattering is the main source
of electron spin relaxation in these samples. This conclusion is also
supported by the theoretical results obtained by Glazov and 
coworkers.~\cite{Glazov03b} In contrast to the theoretically predicted 
non-monotonic behavior with a maximum at T $\approx$ 50K, the experimental 
results suggest that, for this electron density, the D'yakonov-Perel' 
spin lifetime $\tau_{DP}$ grows monotonically with temperature. This is 
probably due to the neglect of phonons in the theoretical modeling. The 
maximum of $\tau_{DP}$ is a consequence of
electron-electron scattering. Additional scattering processes destroy
the maximum. For instance, electron-impurity scattering increases
the spin lifetimes at low temperatures, resulting in a monotonically 
decreasing spin lifetime (see $x\neq 0$ curves
in Fig. \ref{fig6}). The samples in Ref.~\cite{Brand02} are high-quality
quantum wells, where electron-impurity scattering should be negligible. 
The electron density $n=1.86\times10^{11}{\rm cm}^{-2}$ is however rather
high. As a result, the temperature, at which $\tau_{DP}$ is expected
to be maximal, falls in a temperature range, where electron-phonon 
scattering is significant.  We expect a calculation, which takes
electron-electron and electron-phonon scattering into account, to 
produce a monotonically increasing $\tau_{DP}$. 

In this section we illustrated our semiclassical kinetic theory
of spin relaxation by calculating the DP spin lifetime for an
(idealized) quantum well, at temperatures and electron densities,
where electron-electron and electron-impurity scattering dominate.
Electron-electron scattering has been chosen to illustrate the
effects of inelasticity and Pauli blocking. Electron spin lifetimes
due to electron-electron scattering have been also calculated in
Refs.\cite{WuMetiu00,Glazov02,Glazov03a,Glazov03b,Boguslawski80}, using 
however different approaches and focusing mostly on different aspects.
In particular, the non-monotonous temperature and density dependence
has not been addressed until quite recently.~\cite{Brand02,Glazov03b}
In modulation doped quantum wells, spin lifetimes
turn out to be particularly long for electron densities and
temperatures, where the cross-over from the non-degenerate to degenerate
regime occurs. In this regime, many-body effects beyond the Born
approximation are most probably important and should be included in
a more quantitative calculation of the spin relaxation times. We
expect however our main conclusions to be independent of the
particular modeling of the Coulomb interaction.     

\section{Conclusions}

Starting from the full quantum kinetic equations for the
electron Green functions we derived a (semiclassical)
Fokker-Planck equation
for the non-equilibrium spin polarization, assuming small
spin polarizations and soft scattering. The Fokker-Planck
equation conceptualizes the non-equilibrium spin dynamics in terms of a
``test'' spin polarization, comprising a small number of spin
polarized ``test'' electrons, which scatter off an equilibrated bath consisting
of impurites, phonons, and spin-balanced ``field'' electrons.
Because of the scattering, the bath causes for the spin
polarization dynamical friction, diffusion, and relaxation (decay).
We then empolyed a multiple time scale perturbation approach
to separate the fast spin-conserving from the slow spin non-conserving
time evolution. As a result, we extracted from
the Fokker-Planck equation a Bloch equation which controls
the time evolution of the macroscopic ($\vec{k}$-averaged)
spin polarization on the long time scale, where the spin
polarization decays.
Our semiclassical approach accounts for elastic and
inelastic scattering and avoids the ad hoc energy averaging
of on-shell spin relaxation rates. Instead we show that the
weight function is intimately linked to the
``quasi-stationary'' spin polarization, which is the
terminating state of the fast, spin-conserving time evolution
taking place immediately after spin injection.
The diagonal elements of the macroscopic ($\vec{k}$-averaged)
spin relaxation tensor are the spin lifetimes. They are either
given by an energy averaged spin-flip rate (EY process) or
an energy average of a generalized relaxation
time multiplied by a precession rate (DP and VG processes).

The formal development of our approach is based on a generic
model for non-magnetic III-V semiconductors and treats
EY and motional narrowing (DP and VG) spin relaxation processes
on an equal footing. We also allowed for orbital motion of the
electrons in a strong magnetic field, which potentially leads to a
quenching of the motional narrowing-type spin relaxation processes.
The derivation of the Fokker-Planck equation
is independent of dimensionality and, as long as a soft scattering
regime can be identified, also of the scattering processes,
which enter the Fokker-Planck equation in
the form of dynamical friction and diffusion coefficients, which
have to be worked out separately for each scattering process.

To illustrate our formalism we applied it to a quantum well
at low temperatures, where electron-electron and electron-impurity
scattering dominate. We explicitly constructed the
friction, diffusion and angle randomization coefficients characterizing
the (symmetry-adapted) Fokker-Planck equation for that particular situation
and calculated the DP spin lifetime at vanishingly small magnetic field 
as a function of electron density and temperature. We found that for fixed
temperature (density) the density (temperature) dependence is
non-monotonic. Spin lifetimes are particularly long for densities
and temperatures, where the cross-over from a non-degenerate to
a degenerate electron gas occurs. Spin lifetimes in compensated
quantum wells are always longer than in modulation doped
quantum wells with the same electron density. The enhancement of
the spin lifetime is particularly strong for densities and
temperatures where Pauli blocking is most efficient in suppressing
the DP spin lifetime (due to electron-electron scattering) in modulation
doped quantum wells.

Various extensions of our approach are conceivable and constitute
research directions for the future. Semiconductor structures with
structural inversion asymmetry~\cite{SIA} and/or native interface
asymmetry~\cite{Olesberg01} can be studied within our approach by
augmenting the model Hamiltonian by the corresponding spin
off-diagonal Hamiltonian matrix elements. In particular, the role
of the linear collision integral $J_B^{(1)}[f,\delta\vec{S}]$,
which does not affect spin lifetimes in isotropic semiconductors,
should be reinvestigated, e.g., for an asymmetric quantum well
where spin lifetimes can be particularly long because motional
narrowing processes due to bulk and structural inversion asymmetry
can be made to cancel each other.~\cite{Averkiev99,Kainz03} Since
$J_B^{(1)}[f,\delta\vec{S}]$ potentially mixes spin relaxation
channels, it could affect the cancellation. A Fokker-Planck
equation of the form (\ref{Boltzmann0}), perhaps augmented by
additional driving terms, could be the starting point for a
systematic calculation of spin transport coefficients (e.g., spin
diffusion length) for spatially inhomogeneous systems, such as
interfaces or biased heterostructures. Finally, non-linear effects
due to large spin polarizations could be studied either at the
level of the matrix-Boltzmann equation for the electronic density
matrix~\cite{weng1,weng2} or, if the diffusion approximation is used
to simplify the collision terms, at the level of a 
``Fokker-Planck-Landau equation'' for the spin polarization, where
the differential operator describing spin-conserving scattering
events as well as the spin-flip tensor explicitly depend on the
spin polarization and the distribution of the spin polarized electrons.

\section{Acknowledgment} This work was supported by the Los Alamos Laboratory
Directed Research and Development program.

\appendix

\section{Calculation of relaxation rates}

In this appendix we calculate for a quantum well the dynamical friction and
diffusion coefficients $A_i(\vec{k})$ and $B_{ij}(\vec{k})$ taking
electron-electron and electron-impurity scattering into account. As a result,
we obtain the relaxation rates $1/\tau_f(\varepsilon)$,
$1/\tau_d(\varepsilon)$, and $1/\tau_\perp(\varepsilon)$ which
define the differential operator ${\cal D}(\vec{k})$ as well
as the spin-flip tensor ${\bf R}(\vec{k})$.
Since we discuss in this paper only the
DP process quantitatively, which originates from the interplay
of the momentum scattering encoded in the differential operator
${\cal D}(\vec{k})$ and the torque force due to
inversion asymmetry, it suffices to give explicit expressions only for
the differential operator ${\cal D}(\vec{k})$. The derivation of the
spin-flip tensor ${\bf R}(\vec{k})$ proceeds along the same lines.

Within the diffusion approximation, the spin conserving collision 
integral $J_\nu^{(0)}[f,\delta\vec{S}]$ becomes a 
Fokker-Planck differential operator (\ref{diffapp}) with  
dynamical friction and diffusion coefficients, $A^\nu_{i}$ and $B^\nu_{ij}$, 
defined in Eqs. (\ref{Ai}) and (\ref{Bij}), respectively. 
First, we consider electron-electron 
scattering and calculate $A^{ee}_{i}$ and $B^{ee}_{ij}$.
To avoid the cut-off problem at large momentum transfers, which
usually plagues the diffusion approximation to electron-electron scattering,
we keep the full integrands in Eqs. (\ref{Ai}) and (\ref{Bij}), i.e., 
we do not expand the
distribution functions and energies with respect to the momentum
transfer $\vec{q}$. Using the identities  
$f(\vec{k})f(\vec{k}\pm\vec{q})=[f(\vec{k}\pm\vec{q})-f(\vec{k})]
n(\varepsilon(\vec{k})
-\varepsilon(\vec{k}\pm\vec{q}))$ 
and 
$\delta(\varepsilon(\vec{k})+\varepsilon(\vec{k}')-\varepsilon(\vec{k}-
\vec{q})-\varepsilon(\vec{k}'+\vec{q}))=\int d\omega
\delta(\varepsilon(\vec{k})-\varepsilon(\vec{k}-\vec{q})-\omega)
\cdot\delta(\varepsilon(\vec{k}')-\varepsilon(\vec{k}'+\vec{q})+\omega)$,
we rewrite the transition probability (\ref{Wee}) into
\begin{eqnarray}
W^{ee}(\vec{k};\vec{q})=2\int^\infty_{-\infty}d\omega|V(q)|^2
{\rm Im} \chi(q,\omega)[n(-\omega)+f(\vec{k}-\vec{q})]\delta(\varepsilon
(\vec{k})-\varepsilon(\vec{k}-\vec{q})-\omega)~,
\label{AppWee1}
\end{eqnarray}
where we introduced the susceptibility of noninteracting electrons
\begin{eqnarray}
{\rm Im}\chi(q,\omega)=2\pi\sum_{\vec{k}}[f(\vec{k}+\vec{q})-f(\vec{k})]
\delta(\varepsilon(\vec{k}+\vec{q})-\varepsilon(\vec{k})-\omega)~.
\end{eqnarray}
The Coulomb potential $V(q)$ is taken to be statically screened with
the screening length given by the Thomas-Fermi expression. Had we allowed
for dynamical screening, $V(q)\rightarrow V(q,\omega)=V_0(q)/\epsilon(q,\omega)$.
If $\epsilon(q,\omega)$ is approximated by the RPA expression, the resulting
Fokker-Planck equation would be at the level of a quantum analog to the
Lenard-Balescu equation.~\cite{LB} The calculation
of the relaxation rates presented below could be also performed
with this more general expression for the Coulomb matrix element. For
simplicity we present here however only the results for the statically
screened Coulomb potential.

To proceed, we introduce dimensionless quantities, measuring
energies and lengths in scaled Rydbergs and Bohr radii, respectively.
In particular, we use $\tilde{R}_0=R_0/s$,
$\tilde{a}_0=\sqrt{s}a_0$, with
$\tilde{R}_0\tilde{a}_0^2={\hbar}^2/2m_0$ and $e^2=2\sqrt{s}\tilde{R}_0\tilde{a}_0$,
and choose $s$ such that $\tilde{R}_0=1{\rm meV}$. The dimensionless
Fokker-Planck operator has the same form as in
Eq. (\ref{diffapp}) with dynamical friction and diffusion coefficients
given by
\begin{eqnarray}
A^{ee}_i(\vec{k})&=&C^m_{ee}\int d\vec{q}~q_i W^{ee}(\vec{k};\vec{q})~,
\\
B^{ee}_{ij}(\vec{k})&=&\frac{C^m_{ee}}{2}\int d\vec{q}~q_i q_j
W^{ee}(\vec{k}; \vec{q})~,
\end{eqnarray}
with $C^m_{ee}=sm^\ast/\varepsilon_b^2\pi m_0$ and
\begin{eqnarray}
W^{ee}(\vec{k};\vec{q})=\int^\infty_{-\infty}d\omega\frac{F(k^2,
\omega,q)}{(q+q_s)^2}\delta(\vec{k} \cdot \vec{q}-\frac{q^2}{2}-
\frac{\omega}{2})~,
\label{AppWee5}
\end{eqnarray}
where $q_s$ is the Thomas-Fermi screening wave number
and $\varepsilon_b$ is the static dielectric constant.
The function $F(k^2,\omega,q)$ originates from the statistics of the
electron gas and is given by
\begin{eqnarray}
F(k^2,\omega,q)&=&{\rm Im} \tilde{\chi}(q,\omega) N(k,\omega)~,
\\
{\rm Im} \tilde{\chi}(q,\omega)&=&\int d\vec{k} [f(k^2+\omega)-f(k^2)]
\delta(\vec{k} \cdot \vec{q}+\frac{q^2}{2}-\frac{\omega}{2})~,
\\
N(k,\omega)&=&f(k^2-\omega)+n(-\omega)~,
\end{eqnarray}
with $f(x)$ and $n(x)$ the Fermi and Bose functions, respectively.
To calculate $A^{ee}_i(\vec{k})$ and $B^{ee}_{ij}(\vec{k})$ for
a $[001]$ quantum well, we first evaluate the integrals for
a fixed coordinate system in which $\vec{k}=k\hat{e}_x$ and then
rotate to an arbitrary coordinate system.
The result can be cast into the form
\begin{eqnarray}
A^{ee}_i(\vec{k})&=&-\frac{8}{k}G(k)k_i~,
\label{Appfriction}\\
B^{ee}_{ij}(\vec{k})&=&\frac{2}{k}H(k)\delta_{ij}+
\frac{2}{k^3}E(k)k_ik_j~,
\label{Appdiffusion}
\end{eqnarray}
respectively, with three functions defined by
\begin{eqnarray}
G(k)&=&-\frac{C^m_{ee}}{4k}\int^{k^2}_{-\infty}d\omega
\int^{q_{max}}_{q_{min}}dq~\frac{q F(k^2,\omega,q)}{(q+q_s)^2}
\frac{z}{\sqrt{1-z^2}}~,
\\
H(k)&=&\frac{C^m_{ee}}{2}\int^{k^2}_{-\infty}d\omega
\int^{q_{max}}_{q_{min}}dq~\frac{q^2 F(k^2,\omega,q)}{(q+q_s)^2}
\sqrt{1-z^2}~,
\\
E(k)&=&-\frac{C^m_{ee}}{2}\int^{k^2}_{-\infty}d\omega
\int^{q_{max}}_{q_{min}}dq~\frac{q^2 F(k^2,\omega,q)}{(q+q_s)^2}
\frac{1-2z^2}{\sqrt{1-z^2}}~.
\end{eqnarray}
Note that these integrals are well defined. The range of
integration originates from the $\phi$-integration which also
gives rise to the factor involving $z=(q^2+\omega)/2kq$.
With Eqs. (\ref{Appfriction}) and (\ref{Appdiffusion}) and a
transformation to the radial variable $\varepsilon=k^2$,
the spin-conserving part of the dimensionless electron-electron 
collision integral is given by (recall that in two dimensions 
$\delta\vec{S}(\varepsilon,\phi,t)$ contains
the factor $J(\varepsilon)/(2\pi)^2n_s=1/8\pi^2n_s$)
\begin{eqnarray}
J_{ee}^{(0)}[f,\delta\vec{S}]={\cal D}_{ee}
(\vec{k})\delta\vec{S}(\varepsilon,\phi,t)~,
\end{eqnarray}
with
\begin{eqnarray}
{\cal D}_{ee}(\vec{k})=-\frac{\partial}{\partial\varepsilon}
v_{ee}(\varepsilon)+\frac{\partial^2}{\partial\varepsilon^2}
w_{ee}(\varepsilon)-u_{ee}(\varepsilon){\cal L}^2~.
\label{DiffFinal}
\end{eqnarray}                                      
Here we have introduced the total angular momentum operator
in two dimensions, $\hat{\cal L}=-i\partial/\partial\phi$, and the
friction, diffusion and angle randomization coefficients,
$v_{ee}(\varepsilon)$, $w_{ee}(\varepsilon)$, and $u_{ee}(\varepsilon)$, which are
linear combinations of the functions $G(k)$, $H(k)$, and $E(k)$
taken at $k=\sqrt{\varepsilon}$. Specifically, they read:
\begin{eqnarray}
v_{ee}(\varepsilon)&=&-\frac{2C^m_{ee}}{\varepsilon^{1/2}}\int^{\varepsilon}_{-\infty}
d\omega
\int^{q_{max}}_{q_{min}}dq~\frac{\omega F(\varepsilon,\omega,q)}{(q+q_s)^2}
\frac{1}{\sqrt{1-z^2}}~,
\\
w_{ee}(\varepsilon)&=&4C^m_{ee}\sqrt{\varepsilon}\int^{\varepsilon}_{-\infty}
d\omega
\int^{q_{max}}_{q_{min}}dq~\frac{q^2 F(\varepsilon,\omega,q)}{(q+q_s)^2}
\frac{z^2}{\sqrt{1-z^2}}~,
\\
u_{ee}(\varepsilon)&=&\frac{C^m_{ee}}{\varepsilon^{3/2}}\int^{\varepsilon}_{-\infty}
d\omega
\int^{q_{max}}_{q_{min}}dq~\frac{q^2 F(\varepsilon,\omega,q)}{(q+q_s)^2}
\sqrt{1-z^2}~.
\end{eqnarray}
The range of integration depends on the sign of $\omega$,
$q_{max}=q^-_{max}(\varepsilon,\omega)\Theta(-\omega)+
q^+_{max}(\varepsilon,\omega)\Theta(\omega)$
and $q_{min}=q^-_{min}(\varepsilon,\omega)\Theta(-\omega)
+q^+_{min}(\varepsilon,\omega)\Theta(\omega)$, with
\begin{eqnarray}
q^-_{min}(\varepsilon,\omega)&=&-\sqrt{\varepsilon}+\sqrt{
\varepsilon-\omega}~, ~~~\omega\le 0~,
\\
q^-_{max}(\varepsilon,\omega)&=&~~\sqrt{\varepsilon}+\sqrt{
\varepsilon-\omega}~, ~~~\omega\le 0~,
\\
q^+_{min}(\varepsilon,\omega)&=&~~\sqrt{\varepsilon}-\sqrt{
\varepsilon-\omega}~, ~~~0\le \omega\le\varepsilon~,
\\
q^+_{max}(\varepsilon,\omega)&=&~~\sqrt{\varepsilon}+\sqrt{
\varepsilon-\omega}~, ~~~0\le \omega\le\varepsilon~.
\end{eqnarray}
The functions $v_{ee}(\varepsilon)$ and $w_{ee}(\varepsilon)$ are the
dynamical friction and diffusion coefficients for the
spin polarization in $\varepsilon$-space.
They originate from the scattering of the
``test'' electrons comprising the spin polarization with the 
equilibrated, spin-balanced
``field'' electrons. Because the scattering is
inelastic, the ``test'' electrons gain ($\omega<0$) or lose ($\omega>0$)
energy by scattering off ``field'' electrons. The on-shell function
$u_{ee}(\varepsilon)$ describes randomization of the angle $\phi$.
The integrals defining $v_{ee}(\varepsilon)$, $w_{ee}(\varepsilon)$, and
$u_{ee}(\varepsilon)$ have to be done numerically. The singularities are
integrable and Gaussian integration proved to be efficient.
The limiting values are
${\rm lim}_{\varepsilon\rightarrow0}u_{ee}(\varepsilon)=u_0/\varepsilon ,$
${\rm lim}_{\varepsilon\rightarrow0}w_{ee}(\varepsilon)=0 ,$ and
${\rm lim}_{\varepsilon\rightarrow0}v_{ee}(\varepsilon)=v_0$. Moreover,
$w_{ee}(\varepsilon)=v_0\varepsilon$ for $\varepsilon\rightarrow0$, i.e.
$dw_{ee}(0)/d\varepsilon=v_{ee}(0)$, which is essential to guarantee spin conservation
of the differential
operator ${\cal D}_{ee}(\vec{k})$ [cp. Eq. (\ref{spinconservation})]. 
The structure of the
differential operator suggests to write
$v_{ee}=\varepsilon/\tau^{ee}_f$, $w_{ee}=\varepsilon^2/\tau^{ee}_d$, and 
$u_{ee}=1/4\tau^{ee}_\perp$, with
$1/\tau^{ee}_f$, $1/\tau^{ee}_d$, and $1/\tau^{ee}_\perp$ relaxation
rates describing energy relaxation, diffusion, and randomization
of the angle due electron-electron scattering, respectively.

The calculation of the (on-shell) relaxation rate due to electron-impurity
scattering proceeds along the same lines. The starting point is  
Eq. (\ref{CI0}), specialized to electron-impurity scattering, 
that is, with $W^\nu(\vec{k},\vec{q})$ given by  Eq. (\ref{Wei}). 
Going through the same steps as in the case of electron-electron
scattering yields for the spin-conserving part of the dimensionless 
electron-impurity collision integral
(as before $\delta\vec{S}(\varepsilon,\phi,t)$ contains the factor
$1/8\pi^2n_s$)
\begin{eqnarray}
J_{ei}^{(0)}[f,\delta\vec{S}]={\cal D}_{ei}
(\vec{k})\delta\vec{S}(\varepsilon,\phi,t)~,
\end{eqnarray}
where the differential operator is now given by
\begin{eqnarray}
{\cal D}_{ei}(\vec{k})=
-u_{ei}(\varepsilon){\cal L}^2~,
\end{eqnarray}
with
\begin{eqnarray}
u_{ei}(\varepsilon)=
\frac{C^m_{ei}}{\varepsilon^{3/2}}\int^{2\sqrt{\varepsilon}}_0
dq
\frac{q^2}{(q+q_s)^2}
\sqrt{1-z^2}~,
\end{eqnarray}
$C_{ei}^m=s(4\pi m^* n_i\tilde{a}_0^2/\varepsilon_b^2 m_0)$
and $z=q/2\sqrt{\varepsilon}$, where $n_i$ is the sheet
density of the impurities. 
Because of the elasticity of electron-impurity scattering, the
differential operator contains only an on-shell term. The function
$u_{ei}=1/4\tau^{ei}_\perp$ defines the relaxation time due to
electron-impurity scattering. It only causes randomization of the
angle $\phi$. The total spin-conserving collision integral,
taking electron-electron and electron-impurity scattering into
account, is given by $D(\vec{k})=D_{ee}(\vec{k})+D_{ei}(\vec{k})$.  

Similar expressions can be derived for electron-phonon scattering.
For bulk semiconductors the calculation proceeds along the 
same lines with the obvious modifications due to the additional 
angle integration.


\end{document}